\documentclass[aip,jcp,amsmath,amssymb,numerical,preprint,longbibliography]{revtex4-1}

\usepackage{graphicx}
\usepackage{subcaption}
\usepackage[section]{placeins}
\usepackage{booktabs}
\usepackage{cancel}
\usepackage{xcolor}
\newcommand{\Op}[1]{\hat{#1}}
\renewcommand{\vec}[1]{\boldsymbol{#1}}
\newcommand{\bra}[1]{\langle\! #1\! \mid}
\newcommand{\ket}[1]{\mid\! #1\! \rangle}
\newcommand{\bracket}[2]{\langle\! #1\! \mid\! #2\! \rangle}

\AtBeginDocument{
  \heavyrulewidth=.08em
  \lightrulewidth=.05em
  \cmidrulewidth=.03em
  \belowrulesep=.65ex
  \belowbottomsep=0pt
  \aboverulesep=.4ex
  \abovetopsep=0pt
  \cmidrulesep=\doublerulesep
  \cmidrulekern=.5em
  \defaultaddspace=.5em
}

\begin{document}

\title{Assessment of simple and intuitive semiempirical approximations for non-adiabatic coupling vectors in the frame of (LC)-TDDFTB}

\author{Alexander Humeniuk}
\author{Roland Mitri\'{c}}
\email{roland.mitric@uni-wuerzburg.de}
\affiliation{Institut f\"{u}r Physikalische und Theoretische Chemie,
  Julius-Maximilians-Universit\"{a}t W\"{u}rzburg,
  Emil-Fischer-Stra\ss e 42,
  97074 W\"{u}rzburg, Germany}

\begin{abstract}
  In this study,
  two different simple and intuitive semiempirical schemes for computing approximate non-adiabatic coupling vectors (NACVs) between the ground
  and excited electronic states are presented. The first approximation makes use of Mulliken transition charges, while the
  second is based on derivative coupling vectors between localized molecular orbitals. Both approximations lend themselves easily
  to implementation within a whole spectrum of semiempirical quantum-chemical semiempirical methods. Here we present the implementation within the tight-binding DFT and benchmark its performance against analytical TD-DFT NAC vectors for a range of planar
  fluorescent chromophores at the Franck-Condon point.
  The pattern of the atomic NAC vectors is often reproduced, but the relative magnitude and total length of the NAC
  vector are often in serious error. Although quantitative predictions are not possible, these simple and intuitive approximations
  allow to explain, in a qualitative way, trends in the electronic coupling in extended molecular systems and complex materials. 
  In this context, we investigate how the non-adiabatic coupling depends on the delocalization length of an excitation in chromophoric oligomers based on a simple model. Finally, we make general qualitative predictions on the size dependence of the fluorescence quantum yields in extended molecular systems, and illustrate those on the example of triply fused porphyrin tapes with increasing length. 
\end{abstract}

\maketitle

\section{Introduction}
Non-adiabatic coupling vectors (NACVs) play an important role in photochemistry. They describe the coupling between Born-Oppenheimer surfaces due to the nuclear kinetic energy and allow transitions between electronic states in the absence of radiation \cite{baer_book}. They are a vital ingredient in
\begin{itemize}
\item non-adiabatic molecular dynamics simulations (surface hopping on ``on the fly'') \cite{tapavicza2013ab},
\item searching for minimal energy conical intersections \cite{ragazos1992optimization}
\item and predicting non-radiative transition rates and fluorescence quantum yields \cite{valiev2018first}.
\end{itemize}

The brute force method for computing non-adiabatic coupling vectors is numerical differentiation of the wavefunction with respect to the atomic positions, which requires at least $3 N_{\text{atoms}}$ electronic structure calculations. In the context of TD-DFT exact coupling vectors can be obtained analytically \cite{send2010first} in an efficient way, but the implementation of the method is complicated. Having a simple and intuitive approximation for NACVs that may be combined with any semiempirical electronic structure method is therefore highly desirable.

In this article we compare two different semiempirical approximation for calculating non-adiabatic coupling vectors between the ground state and an excited state (usually $S_1$), which have been implemented in the frame of tight-binding DFT \cite{humeniuk2017dftbaby}.
  The first approximation is based on transition charges: In analogy with the transition dipole moment, the NACV is obtained simply from the transition charges and the molecular geometry.
  The second approximation, which has been propounded by Abad et.al. \cite{nacs_approx_MOs}, is based on molecular orbitals: Non-adiabatic couplings between Kohn-Sham orbitals are constructed from gradients of the overlap and Hamiltonian matrix between localized atomic orbitals, which are readily available in tight-binding DFT, since the same quantities are needed for evaluation of the energy gradient.
  Abad et.al. tested their approximation in the vicinity of conical intersections, where the magnitude of the NACVs is largely determined by the small energy gap. At these photochemical funnels the non-adiabatic coupling diverges and the transfer of population between electronic states is usually extremely fast. 
  It remains to be investigate how well the approximation performs when the $S_0-S_1$ energy gap is large such as at the $S_1$ minimum or the Franck-Condon point where the
  length of the NACVs and their orientation relative to the normal modes determines the non-radiative transition rate.
  
   The article is structured as follows: After a brief description of the approximations (in sections \ref{sec:nacs_charges} and \ref{sec:nacs_mos}), we investigate how the non-adiabatic coupling depends on the delocalization length of an excitation in chromophoric oligomers (in section \ref{sec:quantum_yield}). We then graphically compare the direction and magnitude of the approximate NACVs with their exact counterparts for a range of organic molecules with bright $\pi\pi^*$ excitations (in section \ref{sec:comparison}). Finally we make some qualitative predictions of fluorescence quantum yields in porphyrin tapes (section \ref{sec:tapes}).
 
   % Then we turn our attention to fluorescence quantum yields. 
% How the delocalization of the excitation helps to reduce the non-adiabatic coupling. how delocalization of the excitation helps to increase the radiative rate while reducing the non-adiabatic coupling
   
% excitonically coupled linear molecular aggregates (squaraine oligomers)
% fully conjugated porphyrin tapes
% influence of the length of chromophoric oligomers 
% |mu| and |tau|   , quantities relevant for fluorescence quantum yields.
   
\section{Theory}
\subsection{Semiempirical approximations}

The first-order non-adiabatic coupling vector between two electronic Born-Oppenheimer states $m$ and $n$ is
\begin{equation}
  \vec{\tau}_{mn} = \bracket{\Psi_m}{\frac{\partial \Psi_n}{\partial \vec{R}}}. \label{eqn:nacv_grad}
\end{equation}

The coupling vector may be expressed as 

\begin{equation}
    \bracket{\Psi_m}{\frac{\partial \Psi_n}{\partial \vec{R}}} = \frac{\bra{\Psi_m} \frac{\partial \Op{H}}{\partial \vec{R}} \ket{\Psi_n}}{E_n - E_m}
    \label{eqn:nacv_ediff}
\end{equation}

by differentiating the electronic Schr\"{o}dinger equation on both sides with respect to the nuclear coordinates $\vec{R}$ and multiplying by $\bra{\Psi_m}$ for $m \neq n$ and rearranging.

The derivation of this expression requires that $\Op{H} \ket{\Psi_n} = E_n \ket{\Psi_n}$ is satisfied exactly, which is a much stronger statement than just requiring that Schr\"{o}dinger's equation is satisfied after projecting onto a finite basis set ${ \{ \ket{\Phi_i} \}_{i=1,\ldots,N_{\text{basis}}}  }$:

\begin{equation}
 \bra{\Phi_i} \Op{H} \ket{\Psi_n} = \tilde{E}_n \bracket{\Phi_i}{\Psi_n}  \quad \quad \forall i=1,\ldots,N_{\text{basis}}
\end{equation}

Therefore eqn.~(\ref{eqn:nacv_ediff}) is strictly correct only if a complete basis set is used. In finite basis sets additional Pulay terms \cite{pulay1969ab} have to be considered which arise from the dependence of the basis set on the nuclear coordinates.

Nevertheless it is a good starting point for semiempirical approximations.

\subsubsection{\label{sec:nacs_charges} Approximation based on transition charges}

Since the electronic Hamiltonian depends on the nuclear geometry only through the Coulomb attraction between nuclei and electrons,

\begin{equation}
  \Op{V}_{ne} = \sum_A^{\text{atoms}} \sum_i^{\text{electrons}} \frac{-Z_A}{\vert \vec{R}_A - \vec{r}_i \vert},
\end{equation}

the coupling vector ~(\ref{eqn:nacv_ediff}) on atom $A$ simplifies to

\begin{equation}
\vec{\tau}^A_{mn} = \frac{Z_A}{E_n - E_m} \int d\vec{r} \frac{\vec{R}_A - \vec{r}}{\vert \vec{R}_A - \vec{r} \vert^3} \rho_{mn}(\vec{r}), \label{eqn:nacv_ediff_enuc}
\end{equation}

where we have also introduced the transition density matrix
\begin{equation}
\rho_{mn}(\vec{r}) = N \int \ldots \int d\vec{r}_2 \ldots d\vec{r}_N \Psi_m^*(\vec{r},\vec{r}_2,\ldots,\vec{r}_N) \Psi_n(\vec{r},\vec{r}_2,\ldots,\vec{r}_N).
\end{equation}

By partial integration of eqn.~(\ref{eqn:nacv_ediff_enuc}) (see appendix \ref{sec:partial_integration}) the NACV turns into

\begin{equation}
    \vec{\tau}^A_{mn} = \frac{-Z_A}{E_n - E_m} \int d\vec{r} \frac{\nabla \rho_{mn}(\vec{r})}{\vert \vec{R}_A - \vec{r} \vert} \label{eqn:nacv_rho_deriv}.
\end{equation}

This expression is very instructive since it shows that the coupling vector density is proportional to the gradient of the transition density. The largest contribution comes from points where $\vec{r} \approx \vec{R}_A$ due to the singularity of the Coulomb potential. Therefore we can say qualitatively that the non-adiabatic coupling vector on atom $A$ is approximately proportional to the gradient of the transition density around that atom.

To derive a semiempirical approximation for $\tau^A_{mn}$ let us return to eqn.~(\ref{eqn:nacv_ediff_enuc}) and assume that the transition density may be approximated by atomic transition charges (monopoles)
\begin{equation}
\rho_{mn}(\vec{r}) \approx \sum_B q_B \delta(\vec{r} - \vec{R}_B). \label{eqn:rho_monopole_approx}
\end{equation}
where $\delta(\cdot)$ is Dirac's $\delta$-function.
This approximation is frequently employed in semiempirical methods such as tight-binding DFT \cite{koskinen2009density}. The transition charges $q_A$ may be fitted to reproduce the electrostatic potential generated by the transition density (using the CHELPG algorithm) \cite{madjet2006intermolecular} or they may be calculated as Mulliken transition charges from the transition density matrix. % (see \ref{sec:mulliken_transition_charges}).  
Substituting the monopole approximation ~(\ref{eqn:rho_monopole_approx}) into eqn. ~(\ref{eqn:nacv_ediff_enuc}) and using the property of the $\delta$-function, $\int \delta(x-x_0) f(x) dx = f(x_0)$, we get

\begin{equation}
\vec{\tau}^A_{mn} \approx \frac{Z_A}{E_n - E_m} \sum_{B \neq A} q_B \frac{\vec{R}_A - \vec{R}_B}{\vert \vec{R}_A - \vec{R}_B \vert^3}. \label{eqn:nacv_trchg}
\end{equation}

The term where $A = B$ was excluded to avoid dividing by zero. Only valence electrons are usually included in semiempirical calculations. Then the bare nuclear charge $Z_A$ should be replaced by the charge of the atomic core $Z_A^{\text{core}}$ (nucleus and core electrons), for instance in the case of carbon $Z_A^{\text{core}}=4$ instead of $Z_A=6$.

This approximation is completely analogous to how the transition dipole moment is calculated from the transition charges in the frame of TD-DFTB \cite{niehaus2001tight},

\begin{equation}
\vec{\mu}_{mn} \approx \sum_A q_A \vec{R}_A.  \label{eqn:dipole_trchg}
\end{equation}

\begin{figure}[h!]
  \centering
  \includegraphics[width=0.9\textwidth]{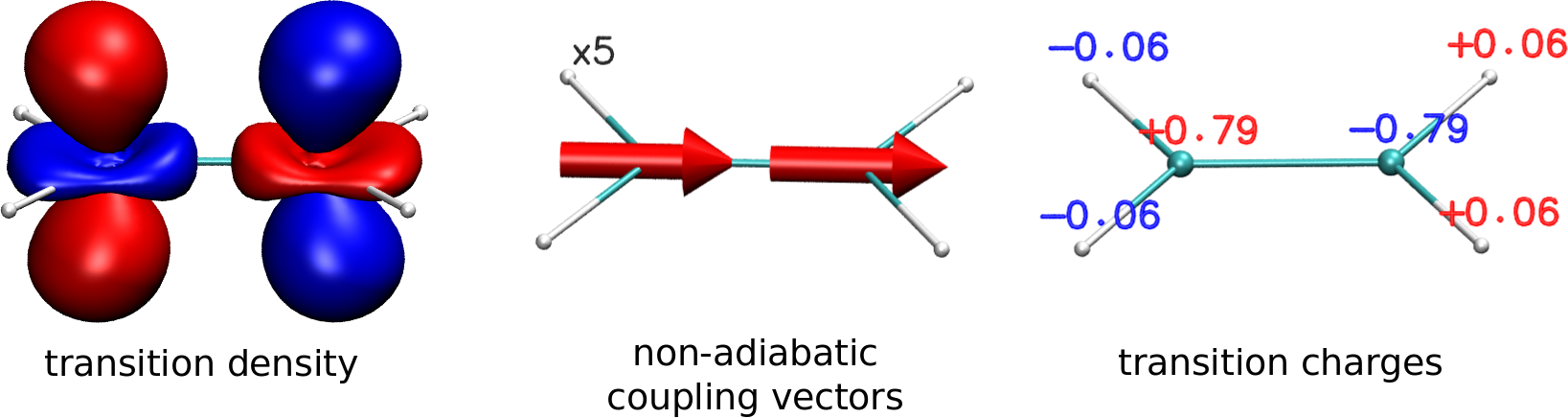}
  \caption{\textbf{Ethene, $\pi\pi^*$ transition }.}
  \label{fig:ethene_nacs_charges_trdensity}
\end{figure}

The simplicity of the derived approximate expressions enables us to make some general statements about the properties of the NACVs. The direction and length of NACVs can be deduced qualitatively by inspecting the transition density or the distribution of the transition charges:
\begin{itemize}
\item Coupling vectors are non-zero only on atoms which take part in an excitation.
\item The coupling vectors point roughly along the direction where the transition density changes most strongly.
  Thus, if there is a node in the transition density between two atoms, the NACV on the atom is perpendicular to the nodal surface. 
\end{itemize}
As a simple example consider the $\pi\pi^*$ excitation in ethene (Fig. \ref{fig:ethene_nacs_charges_trdensity}). The transition charge is positive on one carbon,
negative on the other and almost zero on the hydrogen atoms. Therefore the coupling vectors on the hydrogen atoms are zero. The transition charges change strongly
from $+q$ to $-q$ when moving from one carbon to the other along the C$=$C bond. Therefore the NACVs on the carbons point along this bond.

The approximation fails completely when the transition density cannot be adequately described by monopoles.
For instance in water, the HOMO-LUMO transition, $4a_1 \leftarrow 1b_1$, has lobes of opposite sign below and above the molecular plane.
The gradient of the transition density points perpendicularly to the molecular plane and is orthogonal to all vectors $\vec{R}_A - \vec{R}_B$.
This implies that the coupling vector cannot be represented in the basis of bond vectors.
The Mulliken transition charges are all zero, as is the approximate non-adiabatic coupling vector.
In this case the approximation for the electric transition dipole given in Eq. \ref{eqn:dipole_trchg} is also incorrect.

\subsubsection{\label{sec:nacs_mos} Approximation based on molecular orbitals}

Here we briefly recapitulate how NACVs are calculated in the local-orbital scheme proposed in Ref. \cite{nacs_approx_MOs} using the language of tight-binding DFT (DFTB). 
In DFTB a minimal basis set of valence atomic orbital is used. The molecular orbitals (MO) are linear combinations of these localized basis functions $\ket{\mu}$:
\begin{equation}
\ket{\psi_i} = \sum_{\mu} c_{\mu i} \ket{\mu}
\end{equation}

The coefficients $c_{\mu i}$ for the molecular orbital $i$ are the eigenvector of the Kohn-Sham equation belonging to eigenenergy $\epsilon_i$:
\begin{equation}
\sum_{\nu} \left( H^0_{\mu\nu} - \epsilon_i S_{\mu\nu} \right) c_{\nu i} = 0  \label{eqn:kohn_sham_nonscc}
\end{equation}

Matrix elements of the Kohn-Sham Hamiltonian at the reference density, 
\begin{equation}
  H^0_{\mu\nu} = \bra{\mu} H^{\text{KS}}[\rho_0] \ket{\nu}
\end{equation}
the overlap matrix elements
\begin{equation}
  S_{\mu\nu} = \bracket{\mu}{\nu}
\end{equation}
and their gradients are obtained from Slater-Koster rules\cite{slater_koster}.

With the help of eqn. ~(\ref{eqn:kohn_sham_nonscc}) the authors of Ref. \cite{nacs_approx_MOs} derived an approximate expression for
non-adiabatic coupling vectors between molecular orbitals:
\begin{equation}
\begin{split}
  \vec{d}_{ij}^A & = \bracket{\psi_i}{\frac{\partial \psi_j}{\partial \vec{R}_A}} \\
                 & \approx \frac{1}{\epsilon_i - \epsilon_j} \sum_{\mu,\nu} c^*_{\mu i} c_{\nu j} \left[ - \frac{\partial H^0_{\mu\nu}}{\partial \vec{R}_A} + \frac{\epsilon_i + \epsilon_j}{2} \frac{\partial S_{\mu\nu}}{\partial \vec{R}_A} \right]  \label{eqn:coupling_mos}
\end{split}
\end{equation}

% in tight-binding DFT

In time-dependent density functional theory and its tight-binding version, excited states are represented as linear combinations of singly excited Slater determinants:

\begin{equation}
\ket{\Psi_n} = \sum_{o \in \text{occ}} \sum_{v \in \text{virt}} C_{ov}^{(n)} \ket{\Psi_{o}^v}
\end{equation}

Non-adiabatic coupling vectors between the many-electron ground and excited states are obtained by contraction of the single-particle coupling vectors $\vec{d}_{ij}^A$ with the coefficients $C^{(n)}$:

\begin{equation}
\vec{\tau}^A_{0n} = \bracket{\Psi_0}{\frac{\partial \Psi_n}{\partial \vec{R}_A}} = \sum_{o \in \text{occ}} \sum_{v \in \text{virt}} C_{ov}^{(n)} \vec{d}_{ov}^A  \label{eqn:nacv_mos}
\end{equation}

It is worthwhile to highlight some of approximations made in the above derivation:
(a) Eqn. ~(\ref{eqn:kohn_sham_nonscc}) neglects the dependence of the Hamiltonian on the density ($H[\rho] \approx H[\rho_0]$).
  This allowed to derive the relatively simple expression ~(\ref{eqn:coupling_mos}) for the coupling vectors. However, in our calculations we use expression
  ~(\ref{eqn:coupling_mos}) with MO coefficients obtained from solving the Kohn-Sham equations self-consistently.
(b) In principle, the NACV contains a contribution from changes of the coefficients  $\frac{\partial C^{(n)}_{ov}}{\partial \vec{R}}$, which is neglected.
(c) The exact NACV diverges when the ground and excited state cross ($E_1 \approx E_0$), whereas for the approximate NACV this happens when the HOMO-LUMO gap closes ($\epsilon_{\text{HOMO}} \approx \epsilon_{\text{LUMO}}$). It is well-known that
  HOMO-LUMO gaps are often significantly larger than $S_0-S_1$ excitation energies obtained from TD-DFT due to the mixing of single excitations. This suggests that the approximation will work best when such many-body effects are small so that the $S_0-S_1$ transition predominantly has HOMO $\to$ LUMO character.

Expression ~(\ref{eqn:coupling_mos}) is ideally suited for tight-binding DFT, since the gradients of the matrix elements can be constructed very efficiently at runtime
from precalculated Slater-Koster tables. Since the same quantities are needed for assembling the gradient of the energy,
which is needed in any molecular dynamics (MD) simulation, the computation of the NACVs comes at little additional cost.
 This should be contrasted with the computational cost of an alternative method for computing the non-adiabatic couplings in MD simulations:
 In the surface hopping method \cite{tully1990molecular} the electronic populations depend only on the scalar product between the NACV and the nuclear velocity vector. This scalar can
 be obtained directly from the overlap of the electronic wavefunctions at neighbouring timesteps \cite{mitric2009nonadiabatic} obviating the need for computing the NACVs:
\begin{equation}
\bracket{\Psi_m}{\frac{\partial \Psi_n}{\partial \vec{R}}} \cdot \frac{d\vec{R}}{dt} \approx \frac{1}{\Delta t} \bracket{\Psi_m(t)}{\Psi_n(t+\Delta t)}
\end{equation}
However, since each excited state is a linear combination of Slater determinants, the evaluation of the overlap entails a large number of determinants,
rendering this scheme very expensive for large molecules, unless cutoff thresholds are used for culling determinants which contribute little to the overlap integral.

\subsection{\label{sec:quantum_yield}  Qualitative fluorescence quantum yield}

Approximations ~(\ref{eqn:nacv_trchg}) and ~(\ref{eqn:dipole_trchg}) provide qualitative guidelines on how to tune the electronic wavefunctions for increasing the fluorescence quantum yields. At the moment we will only focus on electronic effects, although vibrational effects can also be very important, as will become clear later on.

If the vibrational wavefunction is neglected, according to Fermi's Golden rule the rates for radiative (spontaneous emission) and non-radiative (internal conversion) decays are proportional to the lengths squared of the transition dipole and non-adiabatic coupling vectors, respectively, between the ground state $S_0$ and the first excited state $S_1$:

\begin{eqnarray}
  k_{\text{rad}} &\propto \vert \vec{\mu}_{01} \vert^2 \\
  k_{\text{IC}} &\propto \vert \vec{\tau}_{01} \vert^2 
\end{eqnarray} 

To increase the fluorescence quantum yield
\begin{equation}
\text{QY} = \frac{1}{1 + \frac{k_{\text{IC}}}{k_{\text{rad}}}}
\end{equation}

$k_{\text{rad}}$ needs to be maximized while $k_{\text{IC}}$ needs to be minimized.

This can be achieved by
\begin{itemize}
\item increasing the length of the transition dipole
\end{itemize}
and/or
\begin{itemize}
\item avoiding conical intersections, where $E_1 = E_0$
\item and reducing the gradient of the transition density.
\end{itemize}

To avoid the crossing of the energy levels of $S_1$ and $S_0$, the geometry should be rigid, so that we can assume there is a stable minimum on $S_1$ and the reorganization energy is small. Then it remains to maximize the length of the transition dipole moment and to minimize the gradient of the transition density. Since it is easier to analyze only one factor, we build a simple model, where the transition dipole is constant and only the length of the NAC vector changes.

\subsubsection{\label{sec:1d_model} 1D model fluorophore}

\begin{figure}[h!]
\centering
\includegraphics[width=1.0\textwidth]{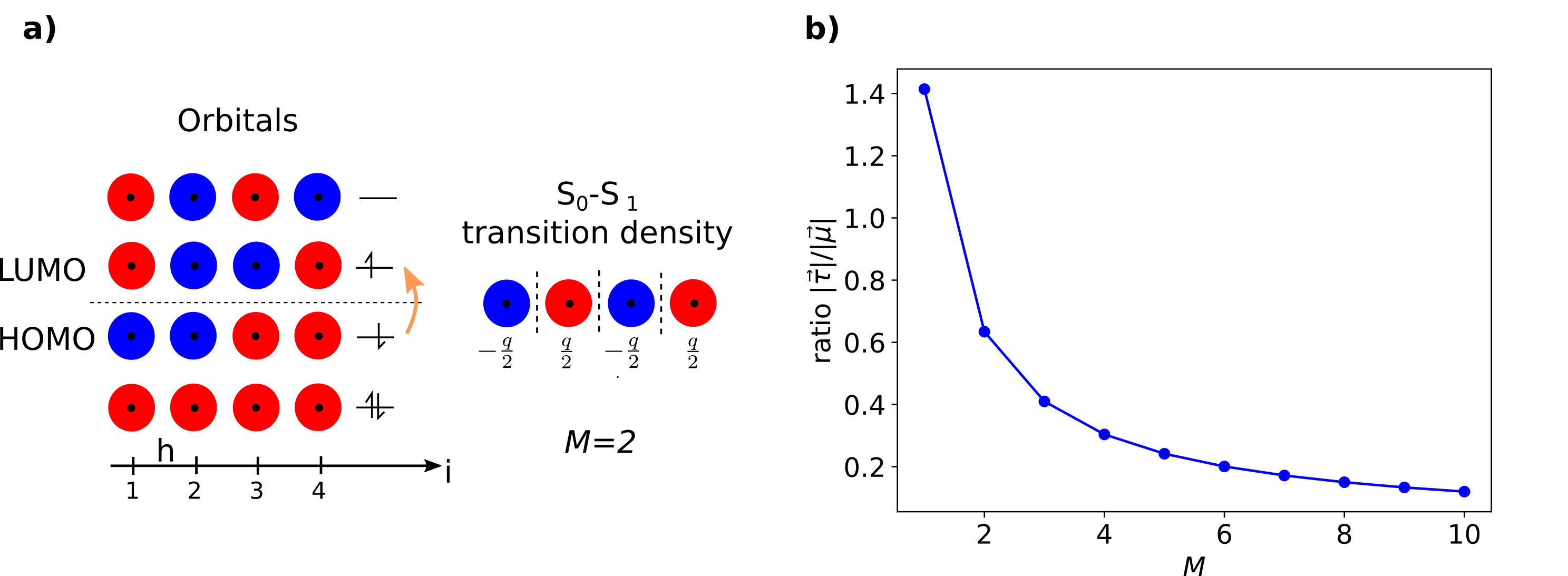}
\caption{1D model fluorophore. \textbf{a)} Linear molecule with 4 atoms. The transition density has nodes between neighbouring atoms.  \textbf{b)} The ratio of the non-adiabatic coupling to the transition dipole moment decreases with the number of atoms $2 M$ taking part in the excitation.}
\label{fig:1d_model_fluorophore}
\end{figure}

Consider a linear molecule (e. g.  a polyene) with $2 M$ atoms on an equidistant grid with spacing $h$ (see Fig. \ref{fig:1d_model_fluorophore}a). For simplicity, we assume that each atom has a single $p_z$ orbital and contributes one electron. The $S_1$ state is a HOMO-LUMO transition. The $S_0-S_1$ transition density has nodes between all atoms, so that the transition charges alternate between positive and negative values. 

The atomic positions and transition charges for atom $i$ are given by
\begin{eqnarray}
  x_i &=& i h  \\
  q_i &=& (-1)^i \frac{q}{M}
\end{eqnarray}

For simplicity we set the nuclear charge to $Z=1$ and the excitation energy to $E_1 - E_0=1$.

The transition dipole moment is independent of the number of atoms (see appendix \ref{sec:dipole_constant}):
\begin{equation}
 \vert \vec{\mu} \vert = \sum_{i=0}^{2M-1} q_i x_i = q h
\end{equation}

The non-adiabatic coupling vector on the $i$-th atom is given by
\begin{equation}
\tau^{(i)} = \frac{q}{M h^2} \sum_{j \neq i}^{2M-1} (-1)^j \frac{i-j}{\vert i-j \vert^3}.
\end{equation}

The length of the total non-adiabatic coupling vector is given by
\begin{equation}
\vert \vec{\tau} \vert = \sqrt{\sum_{i=0}^{2M-1} (\tau^{(i)})^2}.
\end{equation}

and needs to be evaluated numerically. If we plot the length of $\vec{\tau}$ against the number of atoms $2 M$ that participate in the excitation (Fig. \ref{fig:1d_model_fluorophore}b), we see that the rate for internal conversion can be minimized by spreading the transition charge over as many atoms as possible while maintaining the same electric transition dipole moment. Since the transition charges change sign every second atom, the gradient of the transition density can be reduced only if the charges themselves are small. In order to keep the same transition dipole moment, the number of atoms over which the excitation is delocalized needs to be increased.

%\section{Quality of Approximation}
\section{Results}
\subsection{\label{sec:comparison} Comparison between approximate and exact NACVs}

% In TD-DFTB the ordering of the states is not always correct

% In ethene, butadiene and hexatriene the lowest state is the bright one
%   
The two approximations for NACVs are tested for a series organic molecules with bright $\pi\pi^*$ transitions. Many of the selected molecules are fluorescent dyes which have a stable lowest excited singlet state (with the exception of the polyenes).

%\textsl{Computational Details:}
After optimizing the geometries at the AM1 level of theory, the lowest bright excited state was computed with TD-$\omega$B97XD/def2-SVP using Gaussian 16 \cite{g16}. Analytical NACVs were obtained in the frame of TD-DFT \cite{send2010first} via the keyword \textsl{TD=NAC}. These vectors serve as ``exact'' reference values against which the quality of the approximate vectors is measured. Approximate NACVs based on either Mulliken transition charges (according to eqn. ~(\ref{eqn:nacv_trchg})) or localized orbitals (according to eqns. ~(\ref{eqn:coupling_mos}) and ~(\ref{eqn:nacv_mos})) were computed in the frame of long-range corrected tight-binding DFT \cite{humeniuk2017dftbaby}. 
The comparison between the three types of NACVs is presented in a graphical way in Figs. \ref{fig:polyenes_nacs} to \ref{fig:porphyrin_tapes_nacs} below.
The components of the NAC vectors on each atom are shown as little red arrows. Since eigenvalue solvers produce eigenvectors with arbitrary global signs, only the relative orientation of the vectors to each other is important. A sign change in either the bra or the ket wavefunction is equivalent to flipping all vectors simultaneously.
The NACVs were scaled by a factor (which is indicated in the upper left corner) so that the largest vectors in each figure has approximately the same length. We proceed by analyzing the quality of the semiempirical approximations as compared to the exact NACVs for each class of molecules:
 
% polyenes
Trans-polyenes (Fig. \ref{fig:polyenes_nacs}) are the simplest conjugated systems which behave like the linear 1D model discussed above. The lowest bright state ($B_u$) is polarized along the molecular axis. 
The transition density has nodes between neighbouring carbon atoms, which coincide with the positions where the tip of one arrow touches the tail of the next. The individual
arrows become shorter as the number of carbon atoms increases from ethene to hexatriene, reflecting the smearing out of the transition charges over a larger area. 
The  transition charge (TC) approximation overestimates the NACVs by a factor of 5 but gets the orientation of the vectors right. In turn, the localized orbital (LO) approximation
underestimates the NACVs by at least a factor of 10 and fails to predict the orientation. 

% cyanines
The cyanines Cy$N$ (Fig. \ref{fig:cyanine_dyes_nacs}) are fluorescent cationic dyes that consist of a polymethine chain connecting two nitrogens which are part of an indole moiety. Cy3, Cy5 and Cy7 differ by the number of carbon atoms in the bridge, in Cy3B \cite{cooper2004cy3b} the polymethine chain is stabilized against deformation by additional aliphatic rings. In all cyanines the lowest bright excitation is localized on the polymethine chain, and consequently the NACVs are also limited to this region of the molecule. In the polymethine chain the orientation of the arrows alternates as is expected based on the location of the nodes in the transition density.
The LO approximation predicts the position and orientation of the NAC vectors correctly but underestimates their magnitude by a factor of 3. The TC approximation yields NACVs that are spread out too much over non-chromophoric parts of the molecule, such as methyl groups in Cy3-Cy7 or the aliphatic rings in Cy3B. In the polymethine chain the NAC vectors all point in the same direction, but the total magnitude of the NACV is approximately correct.

% squaraines
Dicyanovinyl-substituted squaraines (Fig. \ref{fig:squaraine_dyes_nacs}) \cite{mayerhoffer2013synthesis} are another class of fluorescent dyes.
% cite the article where the first synthesis is reported
In squaraine-O position 3 of the indole moiety is replaced by oxygen, whereas in squaraine-CMe a methyl group is added. The excitation is localized on the central four-membered ring and the adjacent methine groups. The LO approximation reproduces the orientation of the vectors accurately, except for those on the C$\!=\!$O group, which are far too short. The TC approximation places the largest NAC vectors on two opposite carbon atoms in the four-membered ring, although the coupling vectors at these positions should be zero. As with the cyanines a tendency of TC is observed to place large NAC vectors on atoms that are not part of the chromophore.

% polycyclic aromatics
Finally a selection of polycyclic aromatic hydrocarbons is considered in Figs. \ref{fig:aromatic_hydrocarbons_1_nacs} and \ref{fig:aromatic_hydrocarbons_2_nacs}. The couplings were calculated for the lowest excited state. Since the ordering of states can be method-dependent, the symmetry label is given in brackets.

The ring systems give rise to complex patterns in the distribution of NAC vectors.
In fluorene ($B_2$) and phenanthrene ($B_2$) the arrows are arranged in cycles around the outer six-membered rings. This pattern is reproduced by the LO approximation, whereas the TC pattern has no similarity with the exact results. In pyrene ($B_{1u}$), perylene ($B_{1u}$) and rubrene ($B_1$) and relative orientation of the vectors is reproduced correctly both by the LO and the TC approximations, however the relative magnitudes of the vectors differ considerably. The total magnitude of the coupling is severely underestimated by the LO approximation (by a factor of 3-10) and overestimated by the TC approximation (by as much as a factor of 10).
In rubrene the excitation is strictly confined to the tetracene core. Inspite of this, the semiempirical approximations yield large vectors on the adjacent phenyl groups, which are perpendicular to the central tetracene. 

\begin{figure}[h!]
  \centering
  \includegraphics[width=1.0\textwidth]{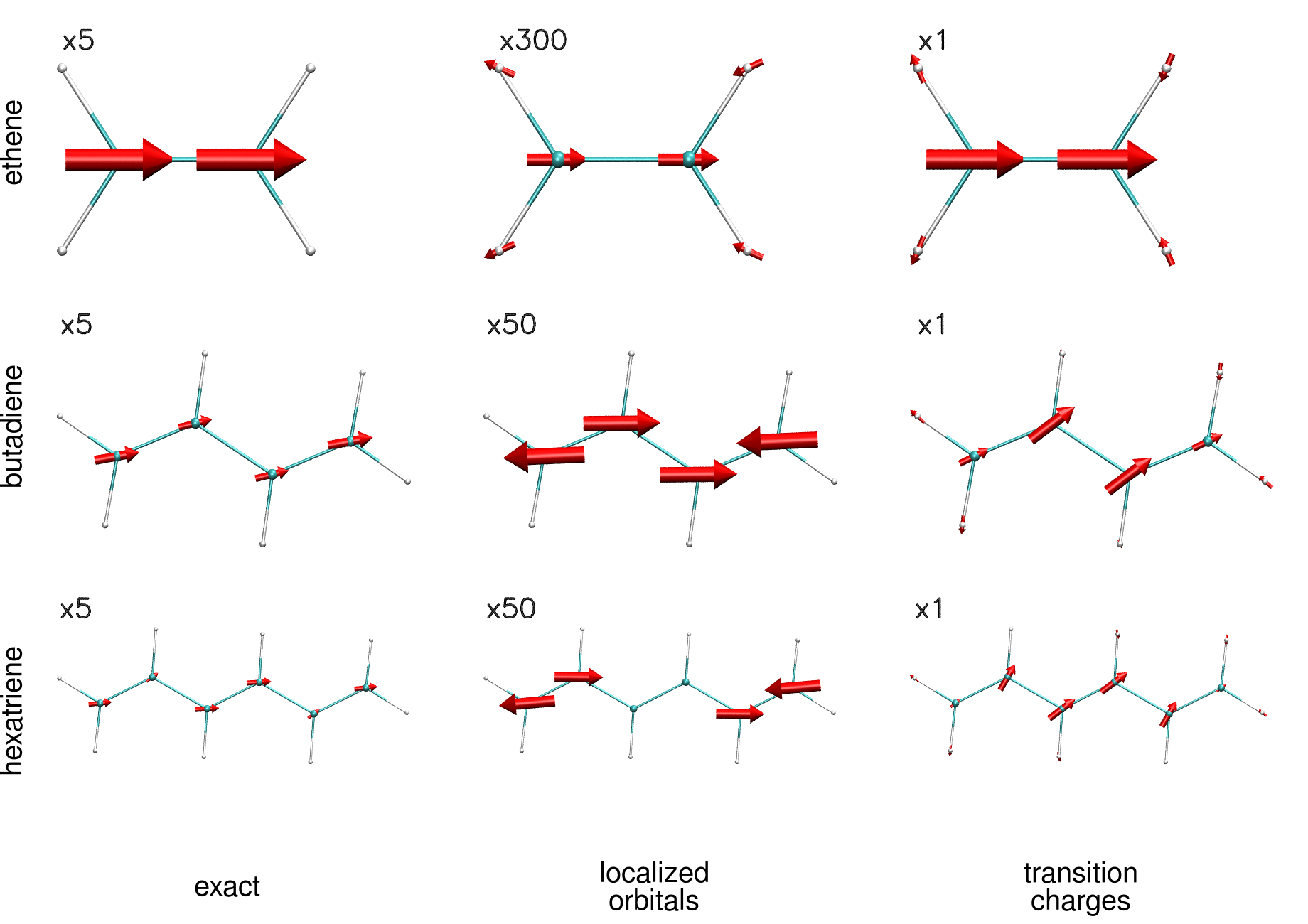}
  \caption{\textbf{Polyenes.}
    Non-adiabatic coupling vectors computed using Furche's analytic method (left)
    and the approximations based on transition charges (middle) or couplings between
    Kohn-Sham orbitals (right).
    The factor by which the vectors where scaled is shown in the upper left corner.}
  \label{fig:polyenes_nacs}
\end{figure}

\begin{figure}[h!]
  \centering
  \includegraphics[width=1.0\textwidth]{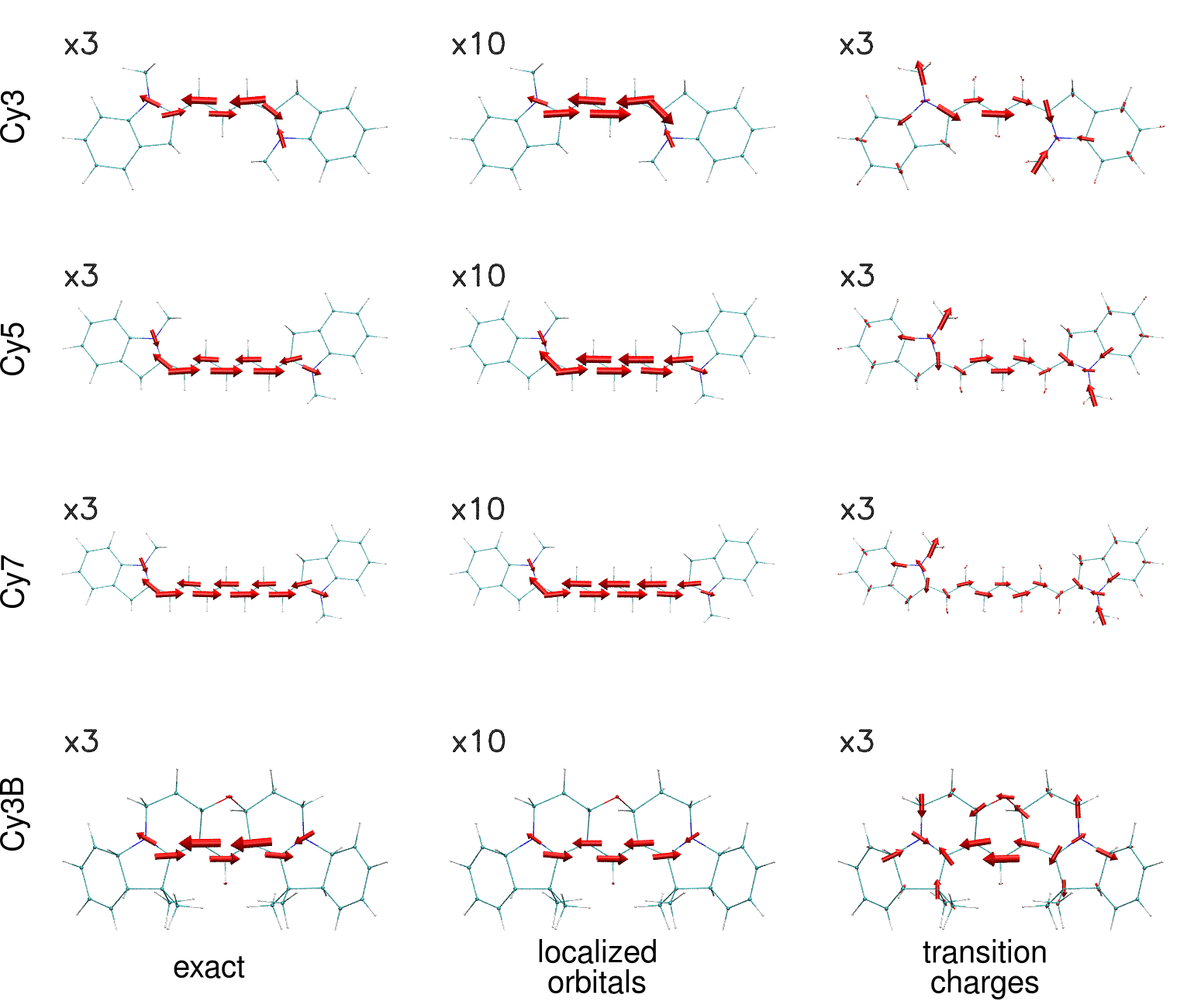}
  \caption{\textbf{Cyanine dyes.}}
  \label{fig:cyanine_dyes_nacs}
\end{figure}

\begin{figure}[h!]
  \centering
  \includegraphics[width=1.0\textwidth]{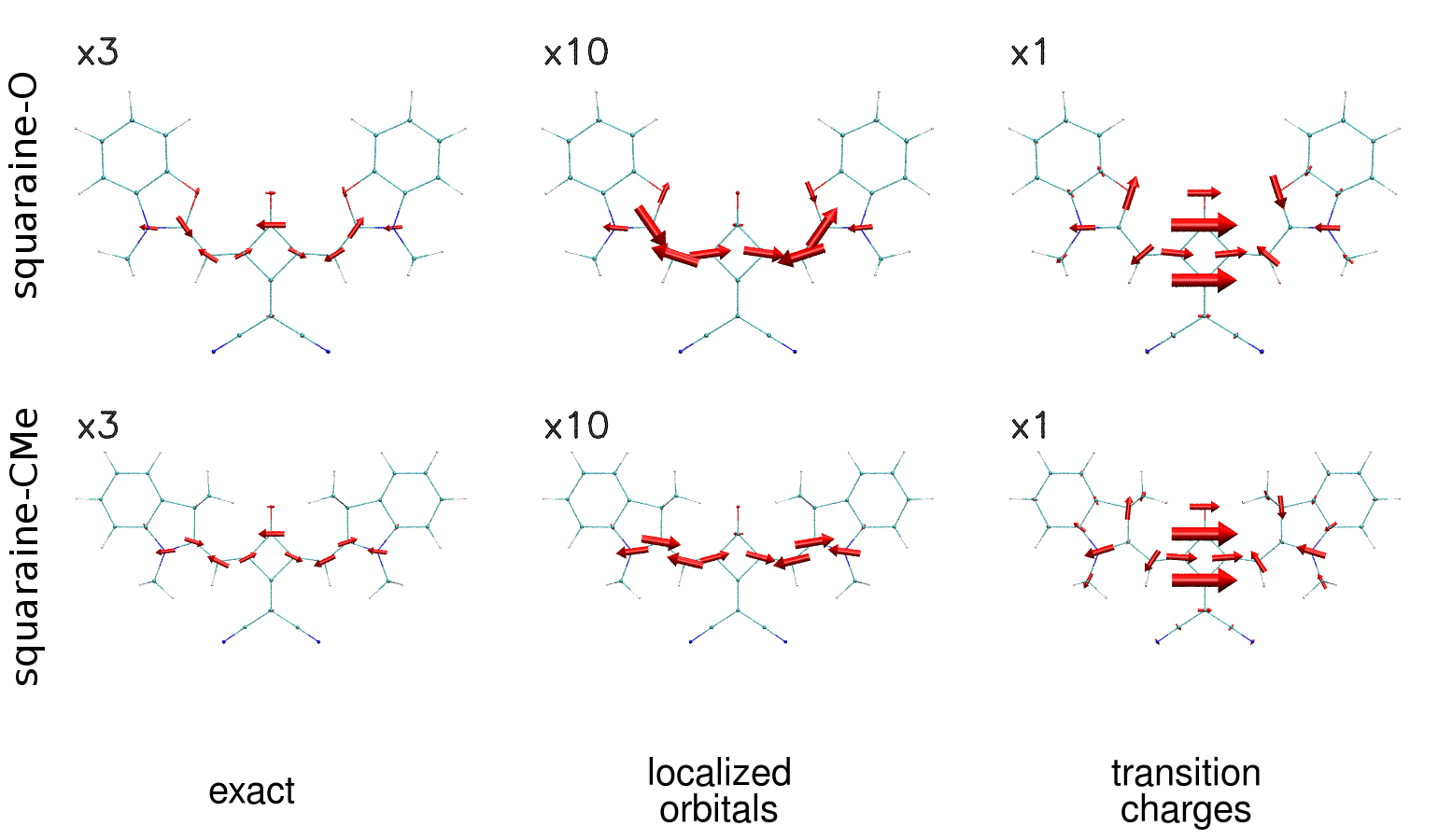}
  \caption{\textbf{Squaraine dyes.}}
  \label{fig:squaraine_dyes_nacs}
\end{figure}

\begin{figure}[h!]
  \centering
  \includegraphics[width=1.0\textwidth]{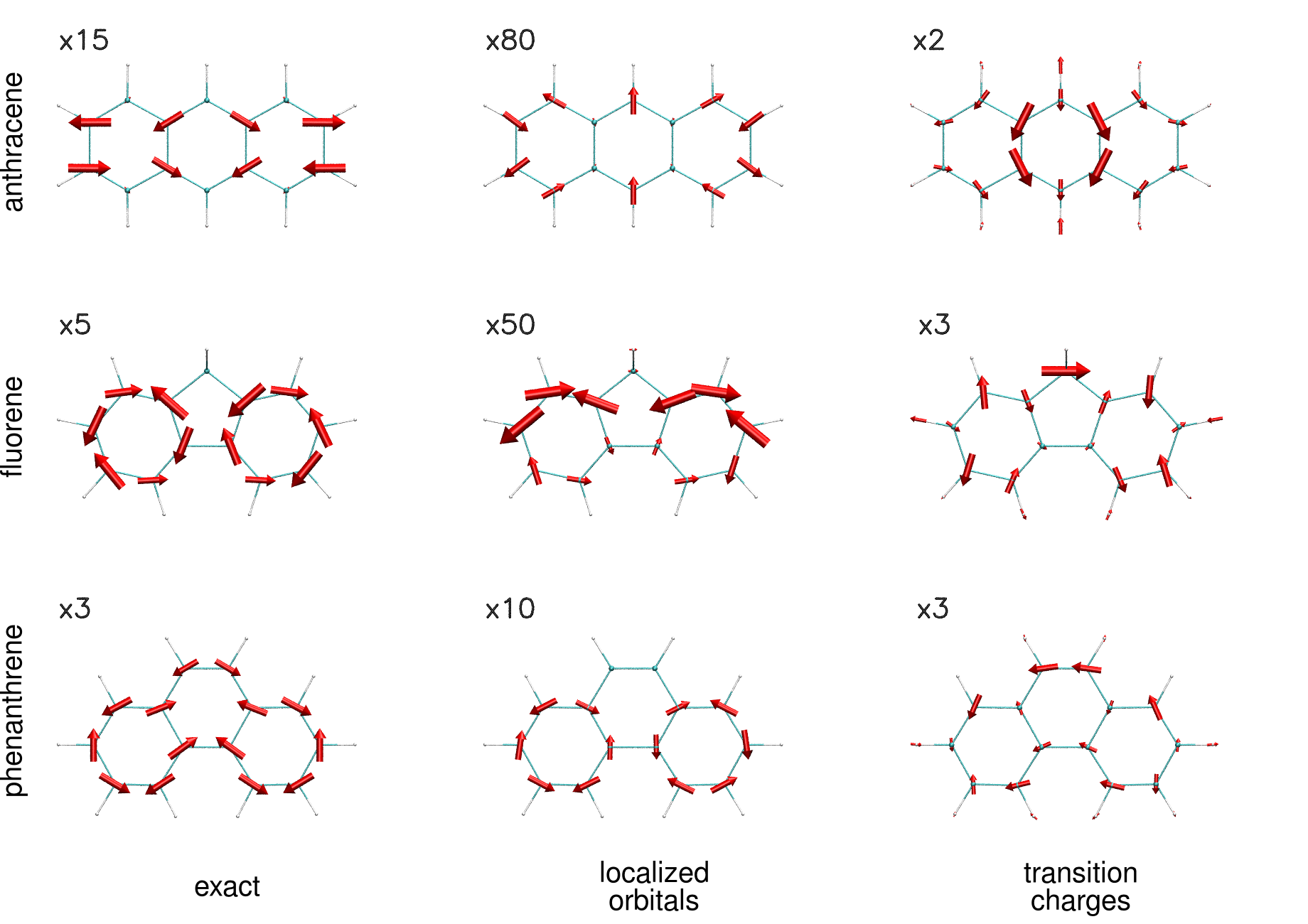}
  \caption{\textbf{Aromatic hydrocarbons 1.}}
  \label{fig:aromatic_hydrocarbons_1_nacs}
\end{figure}

\begin{figure}[h!]
  \centering
  \includegraphics[width=1.0\textwidth]{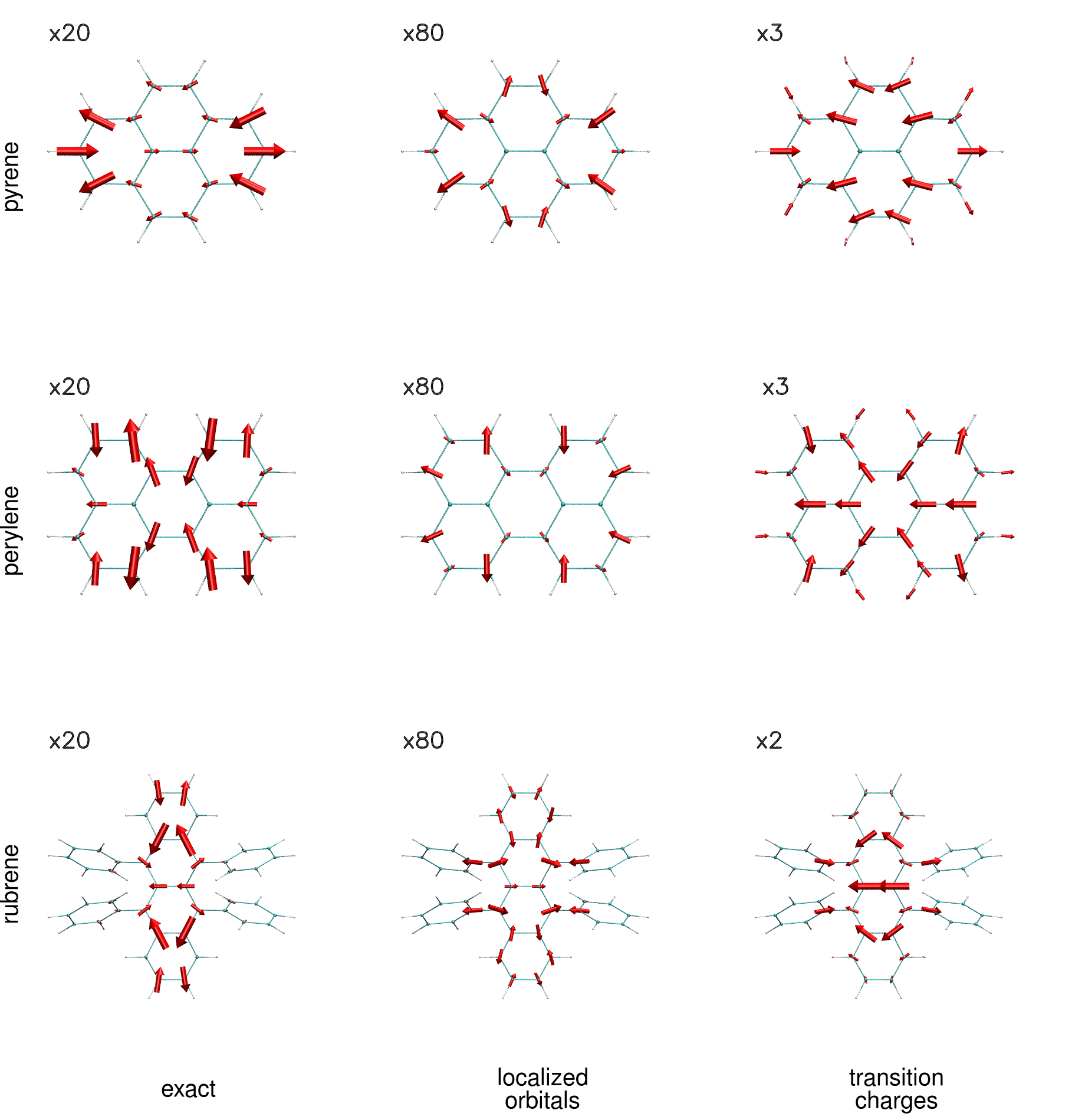}
  \caption{\textbf{Aromatic hydrocarbons 2.}}
  \label{fig:aromatic_hydrocarbons_2_nacs}
\end{figure}

\FloatBarrier

\subsection{\label{sec:tapes} Porphyrin tapes}

\begin{figure}[h!]
  \centering
  \includegraphics[width=0.7\textwidth]{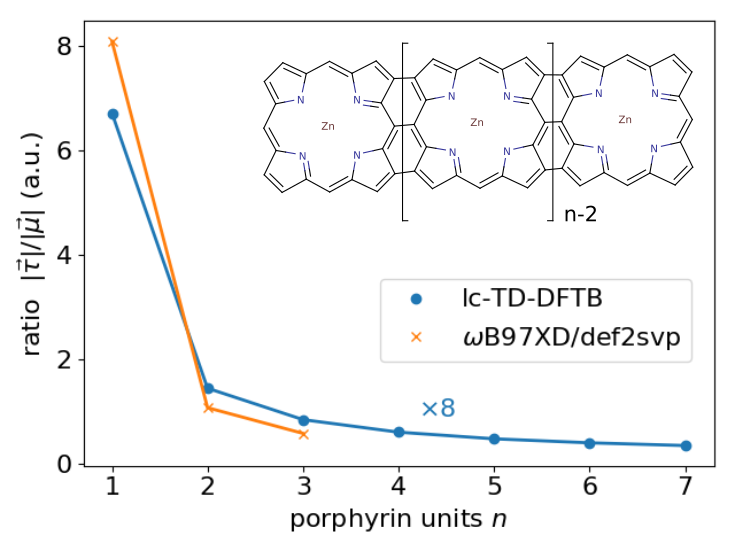}
  \caption{\textbf{Porphyrin tapes.} The ratio of the non-adiabatic coupling vector to the transition dipole moment (in a.u.) is shown as a function of the length $n$ of the porphyrin tape. Observe the similarity with Fig. \ref{fig:1d_model_fluorophore}b. Since NACVs computed with localized orbitals are systematically too low, the semiempirical curve had to be scale by a factor of $8$ to agree with the DFT curve.}
  \label{fig:porphyrin_tapes_nacv_tdip_size_dependence}
\end{figure}

We will now test the predictions of the 1D model from section \ref{sec:1d_model} for the porphyrin tapes that were synthesized by the Tsuda group \cite{tsuda2001fully}. These tapes consist of triply-fused zinc-porphyrins (the structure is shown as an inset in Fig. \ref{fig:porphyrin_tapes_nacv_tdip_size_dependence}). The monomer units are linked through conjugation allowing the electrons to delocalize freely over the entire tape like particles in a box. The delocalization is reflected in the lowering  of the excitation energy far into the infrared with increasing length. At the same time, delocalization of the transition density should also impact the magnitude of the electronic non-adiabatic coupling. 

\begin{figure}[h!]
  \centering
  \includegraphics[width=1.0\textwidth]{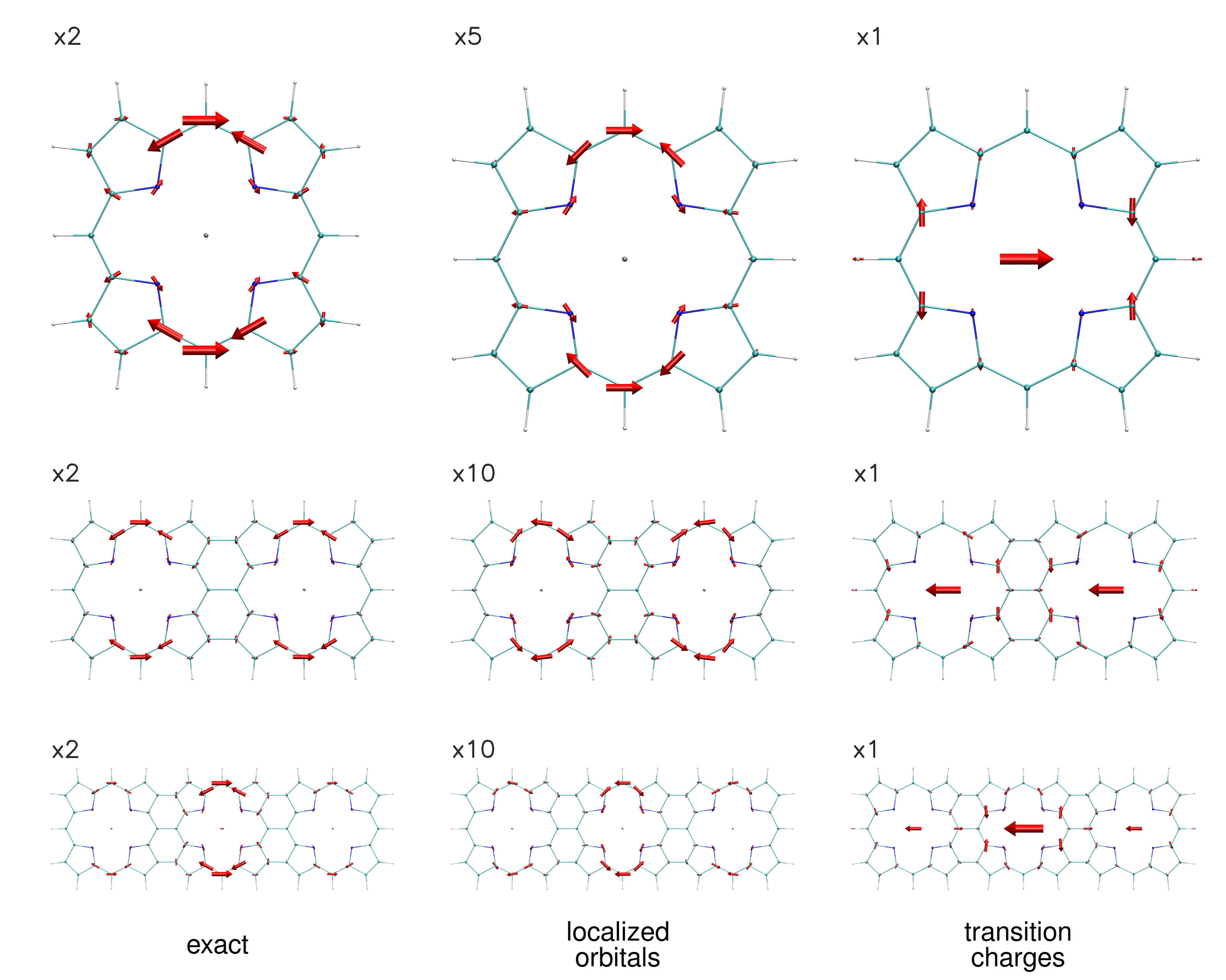}
  \caption{\textbf{Triply-fused porphyrin oligomers.} NACVs between the lowest B$_{1u}$ state and the ground state computed exactly (left) and approximately (center and right).}
  \label{fig:porphyrin_tapes_nacs}
\end{figure}

The transition dipole moments and NACVs
%(based on the local orbital approximation)
were computed with long-range corrected TD-DFTB for the lowest $B_{1u}$ state, which is polarized along the long axis of the tape. For the monomer, dimer and trimer the NACVs are depicted in Fig. \ref{fig:porphyrin_tapes_nacs}. As the conjugation extends over all porphyrin units, the transition dipole moment $\vec{\mu}$ grows approximately linearly with the size of the tape. However, since the transition charges are spread out over a larger area, the non-adiabatic coupling $\vec{\tau}$ grows sublinearly and saturates. The ratio between the lengths of the two vectors is shown in Fig. \ref{fig:porphyrin_tapes_nacv_tdip_size_dependence}. Since the tapes are also very rigid, one can expect that ultrafast internal conversion through conical intersections, which usually requires some local deformation of the geometry, is not the dominant non-radiative decay channel.
Based on this analysis one would expect the long tapes to have extremely high fluorescence quantum yields. 

However, this is not the case. Article \cite{cho2002photophysical} actually shows that the non-radiative rate increases rapidly with the length of the tapes so that the fluorescence is quenched as compared to the monomer. At first this appears to contradict the fact that the ratio of the electronic non-adiabatic coupling to the transition dipole moment, $\tau / \mu$, decreases. However, the reason for the fluorescence quenching in the oligomers is the lower energy gap\cite{tittelbach1995measurements} and the vastly higher density of states. The reduction in the \textsl{electronic} non-adiabatic coupling per porphyrin unit is more than compensated by the increase of accessible final vibrational states.

To verify this explanation we computed the radiative and non-radiative rates for the smallest porphyrin tapes using Fermi's Golden Rule in the harmonic approximation following the steps of Ref. \cite{valiev2018first}.
\footnote{Calculation of rates is based on the following approximations:
 The $S_0$ and $S_1$ potential energy surfaces are harmonic, share the same normal modes and frequencies but have different equilibrium geometries (shifted harmonic oscillators). Frequencies and normal modes are determined from a frequency calculation using $\omega$B97XD/dev2-SVG at the $S_0$ minimum. The Huang-Rhys factors are deduced from the gradient on $S_1$ at the Franck-Condon point. Total rates are obtained by summing over all transitions that start in the initial vibrational ground state on $S_1$ and lead to a final vibrational state on $S_0$. Radiative transitions may lead to any vibrational state lower in energy, while in non-radiative transitions the final vibrational states on $S_0$ have to be approximately isoenergetic with the initial vibrational state on $S_1$. Following Ref. \cite{santoro2007effective} final states are grouped into classes C$_n$ depending on the number $n$ of simultaneously excited modes. The summation is limited to classes $C_0-C_8$, which captures most of the radiative rate and a large fraction of the non-radiative rate. Modes are sorted in decreasing order by Franck-Condon factors and the number of modes from which excitations are allowed, is reduced, until there are no more than $10^8$ elements per class left.}
For the smallest tapes optimizations and frequency calculations on the ground state and the first excited state with $1B_{1u}$ symmetry at the TD-DFT level of theory are still feasible. The resulting rates and quantum yields are listed in table \ref{tbl:porphyrin_tapes_rates}:  

The non-radiative rate jumps by orders of magnitude from the monomer to the dimer and increases further in the trimer. With the non-radiative rate increasing much faster than the radiative rate the quantum yield drops to zero, as observed in experiment. 

 Since the sum over final vibrational states necessarily has to be truncated, the reported non-radiative rates are only a lower limit. Even then it is clear that the non-adiabatic coupling between \textsl{vibrational} states and the sheer density of states is responsible for the fluorescence quenching.

\begin{table}[h!]
  \begin{tabular}{ccccc}
    \toprule
    T$_n$ & $E_{\text{vert}}$ / eV & $k_{\text{rad}}$ (s$^{-1}$) & $k_{\text{nr}}$ (s$^{-1}$) & $QY$ \\
    \midrule
    1 & 2.35 & 9.6e+05 & 3.0e-07 & 1.0e+00 \\
    2 & 1.51 & 1.5e+07 & 8.2e+07 & 1.6e-01 \\
    3 & 1.19 & 3.4e+07 & 3.2e+09 & 1.0e-02 \\
    \bottomrule
  \end{tabular}
  \caption{Dependence of the vertical excitation energy $E_{\text{vert}}$, radiative rate $k_{\text{rad}}$, non-radiative rate $k_{\text{nr}}$ and fluorescence quantum yield $QY = \frac{k_{\text{rad}}}{k_{\text{rad}} + k_{\text{nr}}}$ on the number of porphyrin units $n$ in the triply-fused porphyrin tapes T$_n$.}
  \label{tbl:porphyrin_tapes_rates}
\end{table}

\FloatBarrier

\section{Discussion}

Judging the quality of the NAC vectors by visual inspection can be misleading since it suggests there is more agreement than there actually is.
The symmetry of NAC vectors is related to the symmetry of the excited state. The relative orientation of the vectors in molecules with high symmetry, is therefore largely determined by the irreducible representation. 
In trans-butadiene ($C_{2h}$), for instance, only the relative orientation of two out of four vectors is not already fixed by symmetry. According to TD-DFT these two vectors not related by symmetry should be parallel, but the localized orbital method yields an antiparallel orientation (see Fig. \ref{fig:polyenes_nacs}). The orientation is thus entirely wrong and the magnitude is also wrong by a factor of 10. 

% Magnitude
The localized orbital method tends to underestimate the magnitude of the vectors: In ethene the vectors are too short by a factor of 60, in the cyanine dyes by a factor of 3 and in the porphyrin tapes by a factor of 8. The large error for a system as simple as ethene is surprising. Eqn. ~(\ref{eqn:nacv_rho_deriv}) and Fig. \ref{fig:ethene_nacs_charges_trdensity} showed that the non-adiabatic coupling in the $\pi\pi^*$ state is due to the gradient of the transition density which points along the C-C bond. The transition charge approximation fares a little bit better in predicting the magnitude of the coupling, but it fails in predicting the distribution of the vectors: In the cyanines the excitation is strictly localized on the polyene bridge, but large vectors can be found on two adjacent methyl groups. An extreme example of this are the porphyrin tapes, where the largest vector is placed on the zinc atom, which does not take part in the excitation at all.

% Comparison
Comparison between the two approximations is hindered by the fact that one is derived from eqn. ~(\ref{eqn:nacv_rho_deriv}) (gradient of transition density) and the other from eqn. ~(\ref{eqn:nacv_grad}) (gradient of excited state wavefunction), but the two expressions are only equivalent in the basis set limit, and the minimal valence basis of DFTB is a long way off a complete basis set. 
The LO approximation considers terms that arise because basis functions are attached to the nuclei (Pulay terms) but neglects changes of the excitation coefficients ($\frac{\partial C^{(n)}_{ov}}{\partial \mathbf{R}}$). The TC approximation is independent of the basis set and thus cannot account for Pulay terms. 

% Other sources of errors
Some of the errors relative to TD-DFT might also be due to the tight-binding approximations: Semiempirical transition charges, excitation energies and molecular orbitals, which enter the expressions for the TC and LO approximations,  differ from their ab initio counterparts.
However, those sources of error are of minor importance.
In fact, if we feed our TC approximation with transition charges that were fitted to reproduce the electrostatic potential of the TD-DFT transition density (using the PSPFFT library\cite{budiardja2011parallel} for solving the Poisson equation and the CHELPG algorithm \cite{breneman1990determining}), the resulting vectors are very similar to the tight-binding results.  
The valence basis set employed in DFTB is also not to blame. With a minimal STO-3G basis set the resulting TD-DFT NAC vectors are indistinguishable from the def2-SVP results. 

\section{Conclusion}

Two simple semiempirical approximations for non-adiabatic coupling vectors between excited
singlet states and the ground state were implemented in the frame of (LC)-TDDFTB and compared with TD-DFT coupling vectors as benchmarks
for a set of planar chromophores with bright $S_1$ states.
The TC approximation is based on excitation energies, atom-centered transition charges and geometric information. 
In the LO approximation the coupling between many-body states is calculated from the coupling vectors between
molecular orbitals.

While easy to implement and highly efficient, both approximations are not accurate enough to predict the absolute magnitude of the non-adiabatic coupling vector. In particular the LO approximation underestimates couplings by one order of magnitude. Nevertheless, the region in the molecule where the coupling is large can often be identified. For a series of fused porphyrin tapes the reduction in the electronic coupling per porphyrin unit can be explained by the increasing delocalization of the excitation. As a general rule, spreading transition charges over a larger area reduces the electronic non-adiabatic coupling. This however, does not imply that the fluorescence quantum yield may be increased simply by enlarging the delocalization length, since larger $\pi$-system also have larger nuclear non-adiabatic couplings due to the increased density of states. 

% Possible improvements
The upshot is that quantitative NAC vectors cannot be obtained with these simple approximations. The implementation of analytical coupling vectors in the spirit of Ref. \cite{send2010first} can in principle be adapted to tight-binding DFT in analogy to the analytic gradients \cite{heringer2007analytical} but will require a major effort. The LO approximation is a first step in that direction. Without going to these lengths, the TC approximation might be improved upon by including higher multipoles to represent the transition density more faithfully away from the molecular plane.

\section*{Acknowledgements}
A.H. and R.M. acknowledge funding by the European Research Council (ERC) Consolidator Grant DYNAMO (Grant No. 646737).

\appendix

\iffalse %%%%%%%%%%%%%%%%% commented out %%%%%%%%%%%%%
\section{\label{sec:appendix_nacv_ediff} Detailed derivation of eqn. ~(\ref{eqn:nacv_ediff})}

Starting from the electronic Schr\"{o}dinger equation,

\begin{equation}
\Op{H} \ket{\Psi} = E_n \ket{\Psi_n},
\end{equation}

we differentiate on both sides with respect to the nuclear coordinates $\boldsymbol{R}$, which leads to

\begin{equation}
\frac{\partial \Op{H}}{\partial \vec{R}} \ket{\Psi_n} + \Op{H} \ket{\frac{\partial \Psi_n}{\partial \vec{R}}} = \frac{\partial E_n}{\partial \vec{R}} \ket{\Psi_n} + E_n \ket{\frac{\partial \Psi_n}{\partial \vec{R}}}.
\end{equation}

Multiplying on both sides by $\bra{\Psi_m} \times$ and using the orthogonality of Born-Oppenheimer states belonging to the same geometry, $\bracket{\Psi_m}{\Psi_n} = \delta_{mn}$, and the fact the $\Op{H}$ is Hermitian, $\bra{\Psi_m} \Op{H} = E_m \bra{\Psi_m}$, gives

\begin{equation}
        \bra{\Psi_m} \frac{\partial \Op{H}}{\partial \vec{R}} \ket{\Psi_n}
        + E_m \bracket{\Psi_m}{\frac{\partial \Psi_n}{\partial \vec{R}}}
        =
        \frac{\partial E_n}{\partial \vec{R}} \delta_{mn} + E_n \bracket{\Psi_n}{\frac{\partial \Psi_n}{\partial \vec{R}}}.
\end{equation}

For $m \neq n$ this can be rearranged into

\begin{equation}
    \bracket{\Psi_m}{\frac{\partial \Psi_n}{\partial \vec{R}}} = \frac{\bra{\Psi_m} \frac{\partial \Op{H}}{\partial \vec{R}} \ket{\Psi_n}}{E_n - E_m}
    \label{eqn:nacv_ediff_derivation}
\end{equation}
\fi %%%%%%%%%%%%%%%%%%%%%%%%%%%%%

\section{\label{sec:partial_integration} Detailed derivation of eqn. ~(\ref{eqn:nacv_rho_deriv})}

The single-electron part of the electron-nuclear attraction is
\begin{equation}
 V_{ne}(\vec{r}) = \sum_A \frac{-Z_A}{\vert \vec{R}_A - \vec{r} \vert}.
\end{equation}

\begin{equation}
  \begin{split}
    \vec{\tau}^A_{mn} &= \frac{1}{E_n - E_m} \int d\vec{r} \left(\nabla_A V_{en}(\vec{r}) \right) \rho_{mn}(\vec{r}) \\
    &= \frac{-Z_A}{E_n - E_m} \int d\vec{r} \left(\nabla_A \frac{1}{\vert \vec{R}_A - \vec{r} \vert} \right) \rho_{mn}(\vec{r}) \\
    &= \frac{-Z_A}{E_n - E_m} \int d\vec{r} \left((-\nabla) \frac{1}{\vert \vec{R}_A - \vec{r} \vert} \right) \rho_{mn}(\vec{r})
  \end{split}
\end{equation}

By the partial integration rule $\int_V (\nabla f) g = [f g]_{\partial V} - \int_V f (\nabla g)$ this becomes

\begin{equation}
  \begin{split}
    \vec{\tau}^A_{mn} &= \frac{Z_A}{E_n - E_m} \left\{ \cancel{\left[ \frac{\rho_{mn}(\vec{r})}{\vert \vec{R}_A - \vec{r} \vert} \right]_{\infty}} - \int d\vec{r} \frac{\nabla \rho_{mn}(\vec{r})}{\vert \vec{R}_A - \vec{r} \vert} \right\} \\
    &= - \frac{Z_A}{E_n - E_m} \int d\vec{r} \frac{\nabla \rho_{mn}(\vec{r})}{\vert \vec{R}_A - \vec{r} \vert}
  \end{split}
\end{equation}

%\section{Mulliken transition charges}
%\label{sec:mulliken_transition_charges}
%
%$\alpha$ and $\beta$ enumerate atomic orbitals. The transition density matrix between states $m$ and $n$ in the AO basis is denoted by $P_{\alpha\beta}$. The overlap matrix between atomic orbitals is called $S_{\alpha\beta}$. Then the Mulliken transition charges can be obtained as
%
%\begin{equation}
% q_A = \sum_{\alpha \in A} \sum_{\beta} P_{\alpha,\beta} S_{\alpha,\beta}.
%\end{equation}

\section{\label{sec:dipole_constant} Transition dipole moment in the linear molecule}

The following little calculation shows that the transition dipole moment is a constant, independently of the number of atoms $2 M$:
\begin{equation}
  \begin{split}
    \vert \vec{\mu} \vert &= \sum_{i=0}^{2M-1} q_i x_i = \sum_{i} (-1)^i \frac{q}{M} i h = \ q h \frac{1}{M} \sum_{i=0}^{2M-1} (-1)^i i \\
    &= q h \frac{1}{M} \left\{ \sum_{j=0}^M (-1)^{2j} 2j + \sum_{j=0}^{M-1} (-1)^{2j+1} (2j+1) \right\} \\
    &= qh \frac{1}{M} \left\{ \sum_{j=0}^{M-1} 2j - \sum_{j=0}^{M-1} 2 j + 2 M - \sum_{j=0}^{M-1} 1 \right\} = q h
  \end{split}
\end{equation}

% References
%\nocite{*}
%\bibliographystyle{ieeetr}
\bibliography{references}

%merlin.mbs aipnum4-1.bst 2010-07-25 4.21a (PWD, AO, DPC) hacked
%Control: key (0)
%Control: author (8) initials jnrlst
%Control: editor formatted (1) identically to author
%Control: production of article title (0) allowed
%Control: page (1) range
%Control: year (1) truncated
%Control: production of eprint (0) enabled
\begin{thebibliography}{25}%
\makeatletter
\providecommand \@ifxundefined [1]{%
 \@ifx{#1\undefined}
}%
\providecommand \@ifnum [1]{%
 \ifnum #1\expandafter \@firstoftwo
 \else \expandafter \@secondoftwo
 \fi
}%
\providecommand \@ifx [1]{%
 \ifx #1\expandafter \@firstoftwo
 \else \expandafter \@secondoftwo
 \fi
}%
\providecommand \natexlab [1]{#1}%
\providecommand \enquote  [1]{``#1''}%
\providecommand \bibnamefont  [1]{#1}%
\providecommand \bibfnamefont [1]{#1}%
\providecommand \citenamefont [1]{#1}%
\providecommand \href@noop [0]{\@secondoftwo}%
\providecommand \href [0]{\begingroup \@sanitize@url \@href}%
\providecommand \@href[1]{\@@startlink{#1}\@@href}%
\providecommand \@@href[1]{\endgroup#1\@@endlink}%
\providecommand \@sanitize@url [0]{\catcode `\\12\catcode `\$12\catcode
  `\&12\catcode `\#12\catcode `\^12\catcode `\_12\catcode `\%12\relax}%
\providecommand \@@startlink[1]{}%
\providecommand \@@endlink[0]{}%
\providecommand \url  [0]{\begingroup\@sanitize@url \@url }%
\providecommand \@url [1]{\endgroup\@href {#1}{\urlprefix }}%
\providecommand \urlprefix  [0]{URL }%
\providecommand \Eprint [0]{\href }%
\providecommand \doibase [0]{http://dx.doi.org/}%
\providecommand \selectlanguage [0]{\@gobble}%
\providecommand \bibinfo  [0]{\@secondoftwo}%
\providecommand \bibfield  [0]{\@secondoftwo}%
\providecommand \translation [1]{[#1]}%
\providecommand \BibitemOpen [0]{}%
\providecommand \bibitemStop [0]{}%
\providecommand \bibitemNoStop [0]{.\EOS\space}%
\providecommand \EOS [0]{\spacefactor3000\relax}%
\providecommand \BibitemShut  [1]{\csname bibitem#1\endcsname}%
\let\auto@bib@innerbib\@empty
%</preamble>
\bibitem [{\citenamefont {Baer}(2006)}]{baer_book}%
  \BibitemOpen
  \bibfield  {author} {\bibinfo {author} {\bibfnamefont {M.}~\bibnamefont
  {Baer}},\ }\href@noop {} {\emph {\bibinfo {title} {Beyond
  Born-Oppenheimer}}}\ (\bibinfo  {publisher} {Wiley},\ \bibinfo {year}
  {2006})\BibitemShut {NoStop}%
\bibitem [{\citenamefont {Tapavicza}\ \emph {et~al.}(2013)\citenamefont
  {Tapavicza}, \citenamefont {Bellchambers}, \citenamefont {Vincent},\ and\
  \citenamefont {Furche}}]{tapavicza2013ab}%
  \BibitemOpen
  \bibfield  {author} {\bibinfo {author} {\bibfnamefont {E.}~\bibnamefont
  {Tapavicza}}, \bibinfo {author} {\bibfnamefont {G.~D.}\ \bibnamefont
  {Bellchambers}}, \bibinfo {author} {\bibfnamefont {J.~C.}\ \bibnamefont
  {Vincent}}, \ and\ \bibinfo {author} {\bibfnamefont {F.}~\bibnamefont
  {Furche}},\ }\bibfield  {title} {\enquote {\bibinfo {title} {Ab initio
  non-adiabatic molecular dynamics},}\ }\href@noop {} {\bibfield  {journal}
  {\bibinfo  {journal} {Physical Chemistry Chemical Physics}\ }\textbf
  {\bibinfo {volume} {15}},\ \bibinfo {pages} {18336--18348} (\bibinfo {year}
  {2013})}\BibitemShut {NoStop}%
\bibitem [{\citenamefont {Ragazos}\ \emph {et~al.}(1992)\citenamefont
  {Ragazos}, \citenamefont {Robb}, \citenamefont {Bernardi},\ and\
  \citenamefont {Olivucci}}]{ragazos1992optimization}%
  \BibitemOpen
  \bibfield  {author} {\bibinfo {author} {\bibfnamefont {I.~N.}\ \bibnamefont
  {Ragazos}}, \bibinfo {author} {\bibfnamefont {M.~A.}\ \bibnamefont {Robb}},
  \bibinfo {author} {\bibfnamefont {F.}~\bibnamefont {Bernardi}}, \ and\
  \bibinfo {author} {\bibfnamefont {M.}~\bibnamefont {Olivucci}},\ }\bibfield
  {title} {\enquote {\bibinfo {title} {Optimization and characterization of the
  lowest energy point on a conical intersection using an mc-scf lagrangian},}\
  }\href@noop {} {\bibfield  {journal} {\bibinfo  {journal} {Chemical physics
  letters}\ }\textbf {\bibinfo {volume} {197}},\ \bibinfo {pages} {217--223}
  (\bibinfo {year} {1992})}\BibitemShut {NoStop}%
\bibitem [{\citenamefont {Valiev}\ \emph {et~al.}(2018)\citenamefont {Valiev},
  \citenamefont {Cherepanov}, \citenamefont {Baryshnikov},\ and\ \citenamefont
  {Sundholm}}]{valiev2018first}%
  \BibitemOpen
  \bibfield  {author} {\bibinfo {author} {\bibfnamefont {R.}~\bibnamefont
  {Valiev}}, \bibinfo {author} {\bibfnamefont {V.}~\bibnamefont {Cherepanov}},
  \bibinfo {author} {\bibfnamefont {G.~V.}\ \bibnamefont {Baryshnikov}}, \ and\
  \bibinfo {author} {\bibfnamefont {D.}~\bibnamefont {Sundholm}},\ }\bibfield
  {title} {\enquote {\bibinfo {title} {First-principles method for calculating
  the rate constants of internal-conversion and intersystem-crossing
  transitions},}\ }\href@noop {} {\bibfield  {journal} {\bibinfo  {journal}
  {Physical Chemistry Chemical Physics}\ }\textbf {\bibinfo {volume} {20}},\
  \bibinfo {pages} {6121--6133} (\bibinfo {year} {2018})}\BibitemShut {NoStop}%
\bibitem [{\citenamefont {Send}\ and\ \citenamefont
  {Furche}(2010)}]{send2010first}%
  \BibitemOpen
  \bibfield  {author} {\bibinfo {author} {\bibfnamefont {R.}~\bibnamefont
  {Send}}\ and\ \bibinfo {author} {\bibfnamefont {F.}~\bibnamefont {Furche}},\
  }\bibfield  {title} {\enquote {\bibinfo {title} {First-order nonadiabatic
  couplings from time-dependent hybrid density functional response theory:
  Consistent formalism, implementation, and performance},}\ }\href@noop {}
  {\bibfield  {journal} {\bibinfo  {journal} {The Journal of chemical physics}\
  }\textbf {\bibinfo {volume} {132}},\ \bibinfo {pages} {044107} (\bibinfo
  {year} {2010})}\BibitemShut {NoStop}%
\bibitem [{\citenamefont {Humeniuk}\ and\ \citenamefont
  {Mitri{\'c}}(2017)}]{humeniuk2017dftbaby}%
  \BibitemOpen
  \bibfield  {author} {\bibinfo {author} {\bibfnamefont {A.}~\bibnamefont
  {Humeniuk}}\ and\ \bibinfo {author} {\bibfnamefont {R.}~\bibnamefont
  {Mitri{\'c}}},\ }\bibfield  {title} {\enquote {\bibinfo {title} {Dftbaby: A
  software package for non-adiabatic molecular dynamics simulations based on
  long-range corrected tight-binding td-dft (b)},}\ }\href@noop {} {\bibfield
  {journal} {\bibinfo  {journal} {Computer Physics Communications}\ }\textbf
  {\bibinfo {volume} {221}},\ \bibinfo {pages} {174--202} (\bibinfo {year}
  {2017})}\BibitemShut {NoStop}%
\bibitem [{\citenamefont {Abad}\ \emph {et~al.}(2013)\citenamefont {Abad},
  \citenamefont {Lewis}, \citenamefont {Zoba{\v{c}}}, \citenamefont {Hapala},
  \citenamefont {Jel{\'\i}nek},\ and\ \citenamefont
  {Ortega}}]{nacs_approx_MOs}%
  \BibitemOpen
  \bibfield  {author} {\bibinfo {author} {\bibfnamefont {E.}~\bibnamefont
  {Abad}}, \bibinfo {author} {\bibfnamefont {J.~P.}\ \bibnamefont {Lewis}},
  \bibinfo {author} {\bibfnamefont {V.}~\bibnamefont {Zoba{\v{c}}}}, \bibinfo
  {author} {\bibfnamefont {P.}~\bibnamefont {Hapala}}, \bibinfo {author}
  {\bibfnamefont {P.}~\bibnamefont {Jel{\'\i}nek}}, \ and\ \bibinfo {author}
  {\bibfnamefont {J.}~\bibnamefont {Ortega}},\ }\bibfield  {title} {\enquote
  {\bibinfo {title} {Calculation of non-adiabatic coupling vectors in a
  local-orbital basis set},}\ }\href@noop {} {\bibfield  {journal} {\bibinfo
  {journal} {The Journal of chemical physics}\ }\textbf {\bibinfo {volume}
  {138}},\ \bibinfo {pages} {154106} (\bibinfo {year} {2013})}\BibitemShut
  {NoStop}%
\bibitem [{\citenamefont {Pulay}(1969)}]{pulay1969ab}%
  \BibitemOpen
  \bibfield  {author} {\bibinfo {author} {\bibfnamefont {P.}~\bibnamefont
  {Pulay}},\ }\bibfield  {title} {\enquote {\bibinfo {title} {Ab initio
  calculation of force constants and equilibrium geometries in polyatomic
  molecules: I. theory},}\ }\href@noop {} {\bibfield  {journal} {\bibinfo
  {journal} {Molecular Physics}\ }\textbf {\bibinfo {volume} {17}},\ \bibinfo
  {pages} {197--204} (\bibinfo {year} {1969})}\BibitemShut {NoStop}%
\bibitem [{\citenamefont {Koskinen}\ and\ \citenamefont
  {M{\"a}kinen}(2009)}]{koskinen2009density}%
  \BibitemOpen
  \bibfield  {author} {\bibinfo {author} {\bibfnamefont {P.}~\bibnamefont
  {Koskinen}}\ and\ \bibinfo {author} {\bibfnamefont {V.}~\bibnamefont
  {M{\"a}kinen}},\ }\bibfield  {title} {\enquote {\bibinfo {title}
  {Density-functional tight-binding for beginners},}\ }\href@noop {} {\bibfield
   {journal} {\bibinfo  {journal} {Computational Materials Science}\ }\textbf
  {\bibinfo {volume} {47}},\ \bibinfo {pages} {237--253} (\bibinfo {year}
  {2009})}\BibitemShut {NoStop}%
\bibitem [{\citenamefont {Madjet}, \citenamefont {Abdurahman},\ and\
  \citenamefont {Renger}(2006)}]{madjet2006intermolecular}%
  \BibitemOpen
  \bibfield  {author} {\bibinfo {author} {\bibfnamefont {M.}~\bibnamefont
  {Madjet}}, \bibinfo {author} {\bibfnamefont {A.}~\bibnamefont {Abdurahman}},
  \ and\ \bibinfo {author} {\bibfnamefont {T.}~\bibnamefont {Renger}},\
  }\bibfield  {title} {\enquote {\bibinfo {title} {Intermolecular coulomb
  couplings from ab initio electrostatic potentials: application to optical
  transitions of strongly coupled pigments in photosynthetic antennae and
  reaction centers},}\ }\href@noop {} {\bibfield  {journal} {\bibinfo
  {journal} {The Journal of Physical Chemistry B}\ }\textbf {\bibinfo {volume}
  {110}},\ \bibinfo {pages} {17268--17281} (\bibinfo {year}
  {2006})}\BibitemShut {NoStop}%
\bibitem [{\citenamefont {Niehaus}\ \emph {et~al.}(2001)\citenamefont
  {Niehaus}, \citenamefont {Suhai}, \citenamefont {Della~Sala}, \citenamefont
  {Lugli}, \citenamefont {Elstner}, \citenamefont {Seifert},\ and\
  \citenamefont {Frauenheim}}]{niehaus2001tight}%
  \BibitemOpen
  \bibfield  {author} {\bibinfo {author} {\bibfnamefont {T.~A.}\ \bibnamefont
  {Niehaus}}, \bibinfo {author} {\bibfnamefont {S.}~\bibnamefont {Suhai}},
  \bibinfo {author} {\bibfnamefont {F.}~\bibnamefont {Della~Sala}}, \bibinfo
  {author} {\bibfnamefont {P.}~\bibnamefont {Lugli}}, \bibinfo {author}
  {\bibfnamefont {M.}~\bibnamefont {Elstner}}, \bibinfo {author} {\bibfnamefont
  {G.}~\bibnamefont {Seifert}}, \ and\ \bibinfo {author} {\bibfnamefont
  {T.}~\bibnamefont {Frauenheim}},\ }\bibfield  {title} {\enquote {\bibinfo
  {title} {Tight-binding approach to time-dependent density-functional response
  theory},}\ }\href@noop {} {\bibfield  {journal} {\bibinfo  {journal}
  {Physical Review B}\ }\textbf {\bibinfo {volume} {63}},\ \bibinfo {pages}
  {085108} (\bibinfo {year} {2001})}\BibitemShut {NoStop}%
\bibitem [{\citenamefont {Slater}\ and\ \citenamefont
  {Koster}(1954)}]{slater_koster}%
  \BibitemOpen
  \bibfield  {author} {\bibinfo {author} {\bibfnamefont {C.}~\bibnamefont
  {Slater}}\ and\ \bibinfo {author} {\bibfnamefont {G.}~\bibnamefont
  {Koster}},\ }\bibfield  {title} {\enquote {\bibinfo {title} {Simplified lcao
  method for the periodic potential problem},}\ }\href@noop {} {\bibfield
  {journal} {\bibinfo  {journal} {Phys. Rev.}\ }\textbf {\bibinfo {volume}
  {94}},\ \bibinfo {pages} {1498} (\bibinfo {year} {1954})}\BibitemShut
  {NoStop}%
\bibitem [{\citenamefont {Tully}(1990)}]{tully1990molecular}%
  \BibitemOpen
  \bibfield  {author} {\bibinfo {author} {\bibfnamefont {J.~C.}\ \bibnamefont
  {Tully}},\ }\bibfield  {title} {\enquote {\bibinfo {title} {Molecular
  dynamics with electronic transitions},}\ }\href@noop {} {\bibfield  {journal}
  {\bibinfo  {journal} {The Journal of Chemical Physics}\ }\textbf {\bibinfo
  {volume} {93}},\ \bibinfo {pages} {1061--1071} (\bibinfo {year}
  {1990})}\BibitemShut {NoStop}%
\bibitem [{\citenamefont {Mitric}\ \emph {et~al.}(2009)\citenamefont {Mitric},
  \citenamefont {Werner}, \citenamefont {Wohlgemuth}, \citenamefont {Seifert},\
  and\ \citenamefont {Bonačić-Koutecký}}]{mitric2009nonadiabatic}%
  \BibitemOpen
  \bibfield  {author} {\bibinfo {author} {\bibfnamefont {R.}~\bibnamefont
  {Mitric}}, \bibinfo {author} {\bibfnamefont {U.}~\bibnamefont {Werner}},
  \bibinfo {author} {\bibfnamefont {M.}~\bibnamefont {Wohlgemuth}}, \bibinfo
  {author} {\bibfnamefont {G.}~\bibnamefont {Seifert}}, \ and\ \bibinfo
  {author} {\bibfnamefont {V.}~\bibnamefont {Bonačić-Koutecký}},\
  }\bibfield  {title} {\enquote {\bibinfo {title} {Nonadiabatic dynamics within
  time-dependent density functional tight binding method},}\ }\href@noop {}
  {\bibfield  {journal} {\bibinfo  {journal} {The Journal of Physical Chemistry
  A}\ }\textbf {\bibinfo {volume} {113}},\ \bibinfo {pages} {12700--12705}
  (\bibinfo {year} {2009})}\BibitemShut {NoStop}%
\bibitem [{\citenamefont {Frisch}\ \emph {et~al.}(2016)\citenamefont {Frisch},
  \citenamefont {Trucks}, \citenamefont {Schlegel}, \citenamefont {Scuseria},
  \citenamefont {Robb}, \citenamefont {Cheeseman}, \citenamefont {Scalmani},
  \citenamefont {Barone}, \citenamefont {Petersson}, \citenamefont {Nakatsuji},
  \citenamefont {Li}, \citenamefont {Caricato}, \citenamefont {Marenich},
  \citenamefont {Bloino}, \citenamefont {Janesko}, \citenamefont {Gomperts},
  \citenamefont {Mennucci}, \citenamefont {Hratchian}, \citenamefont {Ortiz},
  \citenamefont {Izmaylov}, \citenamefont {Sonnenberg}, \citenamefont
  {Williams-Young}, \citenamefont {Ding}, \citenamefont {Lipparini},
  \citenamefont {Egidi}, \citenamefont {Goings}, \citenamefont {Peng},
  \citenamefont {Petrone}, \citenamefont {Henderson}, \citenamefont
  {Ranasinghe}, \citenamefont {Zakrzewski}, \citenamefont {Gao}, \citenamefont
  {Rega}, \citenamefont {Zheng}, \citenamefont {Liang}, \citenamefont {Hada},
  \citenamefont {Ehara}, \citenamefont {Toyota}, \citenamefont {Fukuda},
  \citenamefont {Hasegawa}, \citenamefont {Ishida}, \citenamefont {Nakajima},
  \citenamefont {Honda}, \citenamefont {Kitao}, \citenamefont {Nakai},
  \citenamefont {Vreven}, \citenamefont {Throssell}, \citenamefont
  {Montgomery}, \citenamefont {Peralta}, \citenamefont {Ogliaro}, \citenamefont
  {Bearpark}, \citenamefont {Heyd}, \citenamefont {Brothers}, \citenamefont
  {Kudin}, \citenamefont {Staroverov}, \citenamefont {Keith}, \citenamefont
  {Kobayashi}, \citenamefont {Normand}, \citenamefont {Raghavachari},
  \citenamefont {Rendell}, \citenamefont {Burant}, \citenamefont {Iyengar},
  \citenamefont {Tomasi}, \citenamefont {Cossi}, \citenamefont {Millam},
  \citenamefont {Klene}, \citenamefont {Adamo}, \citenamefont {Cammi},
  \citenamefont {Ochterski}, \citenamefont {Martin}, \citenamefont {Morokuma},
  \citenamefont {Farkas}, \citenamefont {Foresman},\ and\ \citenamefont
  {Fox}}]{g16}%
  \BibitemOpen
  \bibfield  {author} {\bibinfo {author} {\bibfnamefont {M.~J.}\ \bibnamefont
  {Frisch}}, \bibinfo {author} {\bibfnamefont {G.~W.}\ \bibnamefont {Trucks}},
  \bibinfo {author} {\bibfnamefont {H.~B.}\ \bibnamefont {Schlegel}}, \bibinfo
  {author} {\bibfnamefont {G.~E.}\ \bibnamefont {Scuseria}}, \bibinfo {author}
  {\bibfnamefont {M.~A.}\ \bibnamefont {Robb}}, \bibinfo {author}
  {\bibfnamefont {J.~R.}\ \bibnamefont {Cheeseman}}, \bibinfo {author}
  {\bibfnamefont {G.}~\bibnamefont {Scalmani}}, \bibinfo {author}
  {\bibfnamefont {V.}~\bibnamefont {Barone}}, \bibinfo {author} {\bibfnamefont
  {G.~A.}\ \bibnamefont {Petersson}}, \bibinfo {author} {\bibfnamefont
  {H.}~\bibnamefont {Nakatsuji}}, \bibinfo {author} {\bibfnamefont
  {X.}~\bibnamefont {Li}}, \bibinfo {author} {\bibfnamefont {M.}~\bibnamefont
  {Caricato}}, \bibinfo {author} {\bibfnamefont {A.~V.}\ \bibnamefont
  {Marenich}}, \bibinfo {author} {\bibfnamefont {J.}~\bibnamefont {Bloino}},
  \bibinfo {author} {\bibfnamefont {B.~G.}\ \bibnamefont {Janesko}}, \bibinfo
  {author} {\bibfnamefont {R.}~\bibnamefont {Gomperts}}, \bibinfo {author}
  {\bibfnamefont {B.}~\bibnamefont {Mennucci}}, \bibinfo {author}
  {\bibfnamefont {H.~P.}\ \bibnamefont {Hratchian}}, \bibinfo {author}
  {\bibfnamefont {J.~V.}\ \bibnamefont {Ortiz}}, \bibinfo {author}
  {\bibfnamefont {A.~F.}\ \bibnamefont {Izmaylov}}, \bibinfo {author}
  {\bibfnamefont {J.~L.}\ \bibnamefont {Sonnenberg}}, \bibinfo {author}
  {\bibfnamefont {D.}~\bibnamefont {Williams-Young}}, \bibinfo {author}
  {\bibfnamefont {F.}~\bibnamefont {Ding}}, \bibinfo {author} {\bibfnamefont
  {F.}~\bibnamefont {Lipparini}}, \bibinfo {author} {\bibfnamefont
  {F.}~\bibnamefont {Egidi}}, \bibinfo {author} {\bibfnamefont
  {J.}~\bibnamefont {Goings}}, \bibinfo {author} {\bibfnamefont
  {B.}~\bibnamefont {Peng}}, \bibinfo {author} {\bibfnamefont {A.}~\bibnamefont
  {Petrone}}, \bibinfo {author} {\bibfnamefont {T.}~\bibnamefont {Henderson}},
  \bibinfo {author} {\bibfnamefont {D.}~\bibnamefont {Ranasinghe}}, \bibinfo
  {author} {\bibfnamefont {V.~G.}\ \bibnamefont {Zakrzewski}}, \bibinfo
  {author} {\bibfnamefont {J.}~\bibnamefont {Gao}}, \bibinfo {author}
  {\bibfnamefont {N.}~\bibnamefont {Rega}}, \bibinfo {author} {\bibfnamefont
  {G.}~\bibnamefont {Zheng}}, \bibinfo {author} {\bibfnamefont
  {W.}~\bibnamefont {Liang}}, \bibinfo {author} {\bibfnamefont
  {M.}~\bibnamefont {Hada}}, \bibinfo {author} {\bibfnamefont {M.}~\bibnamefont
  {Ehara}}, \bibinfo {author} {\bibfnamefont {K.}~\bibnamefont {Toyota}},
  \bibinfo {author} {\bibfnamefont {R.}~\bibnamefont {Fukuda}}, \bibinfo
  {author} {\bibfnamefont {J.}~\bibnamefont {Hasegawa}}, \bibinfo {author}
  {\bibfnamefont {M.}~\bibnamefont {Ishida}}, \bibinfo {author} {\bibfnamefont
  {T.}~\bibnamefont {Nakajima}}, \bibinfo {author} {\bibfnamefont
  {Y.}~\bibnamefont {Honda}}, \bibinfo {author} {\bibfnamefont
  {O.}~\bibnamefont {Kitao}}, \bibinfo {author} {\bibfnamefont
  {H.}~\bibnamefont {Nakai}}, \bibinfo {author} {\bibfnamefont
  {T.}~\bibnamefont {Vreven}}, \bibinfo {author} {\bibfnamefont
  {K.}~\bibnamefont {Throssell}}, \bibinfo {author} {\bibfnamefont {J.~A.}\
  \bibnamefont {Montgomery}, \bibfnamefont {{Jr.}}}, \bibinfo {author}
  {\bibfnamefont {J.~E.}\ \bibnamefont {Peralta}}, \bibinfo {author}
  {\bibfnamefont {F.}~\bibnamefont {Ogliaro}}, \bibinfo {author} {\bibfnamefont
  {M.~J.}\ \bibnamefont {Bearpark}}, \bibinfo {author} {\bibfnamefont {J.~J.}\
  \bibnamefont {Heyd}}, \bibinfo {author} {\bibfnamefont {E.~N.}\ \bibnamefont
  {Brothers}}, \bibinfo {author} {\bibfnamefont {K.~N.}\ \bibnamefont {Kudin}},
  \bibinfo {author} {\bibfnamefont {V.~N.}\ \bibnamefont {Staroverov}},
  \bibinfo {author} {\bibfnamefont {T.~A.}\ \bibnamefont {Keith}}, \bibinfo
  {author} {\bibfnamefont {R.}~\bibnamefont {Kobayashi}}, \bibinfo {author}
  {\bibfnamefont {J.}~\bibnamefont {Normand}}, \bibinfo {author} {\bibfnamefont
  {K.}~\bibnamefont {Raghavachari}}, \bibinfo {author} {\bibfnamefont {A.~P.}\
  \bibnamefont {Rendell}}, \bibinfo {author} {\bibfnamefont {J.~C.}\
  \bibnamefont {Burant}}, \bibinfo {author} {\bibfnamefont {S.~S.}\
  \bibnamefont {Iyengar}}, \bibinfo {author} {\bibfnamefont {J.}~\bibnamefont
  {Tomasi}}, \bibinfo {author} {\bibfnamefont {M.}~\bibnamefont {Cossi}},
  \bibinfo {author} {\bibfnamefont {J.~M.}\ \bibnamefont {Millam}}, \bibinfo
  {author} {\bibfnamefont {M.}~\bibnamefont {Klene}}, \bibinfo {author}
  {\bibfnamefont {C.}~\bibnamefont {Adamo}}, \bibinfo {author} {\bibfnamefont
  {R.}~\bibnamefont {Cammi}}, \bibinfo {author} {\bibfnamefont {J.~W.}\
  \bibnamefont {Ochterski}}, \bibinfo {author} {\bibfnamefont {R.~L.}\
  \bibnamefont {Martin}}, \bibinfo {author} {\bibfnamefont {K.}~\bibnamefont
  {Morokuma}}, \bibinfo {author} {\bibfnamefont {O.}~\bibnamefont {Farkas}},
  \bibinfo {author} {\bibfnamefont {J.~B.}\ \bibnamefont {Foresman}}, \ and\
  \bibinfo {author} {\bibfnamefont {D.~J.}\ \bibnamefont {Fox}},\ }\href@noop
  {} {\enquote {\bibinfo {title} {Gaussian˜16 {R}evision {B}.01},}\ }
  (\bibinfo {year} {2016}),\ \bibinfo {note} {gaussian Inc. Wallingford
  CT}\BibitemShut {NoStop}%
\bibitem [{\citenamefont {Cooper}\ \emph {et~al.}(2004)\citenamefont {Cooper},
  \citenamefont {Ebner}, \citenamefont {Briggs}, \citenamefont {Burrows},
  \citenamefont {Gardner}, \citenamefont {Richardson},\ and\ \citenamefont
  {West}}]{cooper2004cy3b}%
  \BibitemOpen
  \bibfield  {author} {\bibinfo {author} {\bibfnamefont {M.}~\bibnamefont
  {Cooper}}, \bibinfo {author} {\bibfnamefont {A.}~\bibnamefont {Ebner}},
  \bibinfo {author} {\bibfnamefont {M.}~\bibnamefont {Briggs}}, \bibinfo
  {author} {\bibfnamefont {M.}~\bibnamefont {Burrows}}, \bibinfo {author}
  {\bibfnamefont {N.}~\bibnamefont {Gardner}}, \bibinfo {author} {\bibfnamefont
  {R.}~\bibnamefont {Richardson}}, \ and\ \bibinfo {author} {\bibfnamefont
  {R.}~\bibnamefont {West}},\ }\bibfield  {title} {\enquote {\bibinfo {title}
  {Cy3b: improving the performance of cyanine dyes},}\ }\href@noop {}
  {\bibfield  {journal} {\bibinfo  {journal} {Journal of fluorescence}\
  }\textbf {\bibinfo {volume} {14}},\ \bibinfo {pages} {145--150} (\bibinfo
  {year} {2004})}\BibitemShut {NoStop}%
\bibitem [{\citenamefont {Mayerh{\"o}ffer}\ \emph {et~al.}(2013)\citenamefont
  {Mayerh{\"o}ffer}, \citenamefont {Gs{\"a}nger}, \citenamefont {Stolte},
  \citenamefont {Fimmel},\ and\ \citenamefont
  {W{\"u}rthner}}]{mayerhoffer2013synthesis}%
  \BibitemOpen
  \bibfield  {author} {\bibinfo {author} {\bibfnamefont {U.}~\bibnamefont
  {Mayerh{\"o}ffer}}, \bibinfo {author} {\bibfnamefont {M.}~\bibnamefont
  {Gs{\"a}nger}}, \bibinfo {author} {\bibfnamefont {M.}~\bibnamefont {Stolte}},
  \bibinfo {author} {\bibfnamefont {B.}~\bibnamefont {Fimmel}}, \ and\ \bibinfo
  {author} {\bibfnamefont {F.}~\bibnamefont {W{\"u}rthner}},\ }\bibfield
  {title} {\enquote {\bibinfo {title} {Synthesis and molecular properties of
  acceptor-substituted squaraine dyes},}\ }\href@noop {} {\bibfield  {journal}
  {\bibinfo  {journal} {Chemistry--A European Journal}\ }\textbf {\bibinfo
  {volume} {19}},\ \bibinfo {pages} {218--232} (\bibinfo {year}
  {2013})}\BibitemShut {NoStop}%
\bibitem [{\citenamefont {Tsuda}\ and\ \citenamefont
  {Osuka}(2001)}]{tsuda2001fully}%
  \BibitemOpen
  \bibfield  {author} {\bibinfo {author} {\bibfnamefont {A.}~\bibnamefont
  {Tsuda}}\ and\ \bibinfo {author} {\bibfnamefont {A.}~\bibnamefont {Osuka}},\
  }\bibfield  {title} {\enquote {\bibinfo {title} {Fully conjugated porphyrin
  tapes with electronic absorption bands that reach into infrared},}\
  }\href@noop {} {\bibfield  {journal} {\bibinfo  {journal} {Science}\ }\textbf
  {\bibinfo {volume} {293}},\ \bibinfo {pages} {79--82} (\bibinfo {year}
  {2001})}\BibitemShut {NoStop}%
\bibitem [{\citenamefont {Cho}\ \emph {et~al.}(2002)\citenamefont {Cho},
  \citenamefont {Jeong}, \citenamefont {Cho}, \citenamefont {Kim},
  \citenamefont {Matsuzaki}, \citenamefont {Tanaka}, \citenamefont {Tsuda},\
  and\ \citenamefont {Osuka}}]{cho2002photophysical}%
  \BibitemOpen
  \bibfield  {author} {\bibinfo {author} {\bibfnamefont {H.~S.}\ \bibnamefont
  {Cho}}, \bibinfo {author} {\bibfnamefont {D.~H.}\ \bibnamefont {Jeong}},
  \bibinfo {author} {\bibfnamefont {S.}~\bibnamefont {Cho}}, \bibinfo {author}
  {\bibfnamefont {D.}~\bibnamefont {Kim}}, \bibinfo {author} {\bibfnamefont
  {Y.}~\bibnamefont {Matsuzaki}}, \bibinfo {author} {\bibfnamefont
  {K.}~\bibnamefont {Tanaka}}, \bibinfo {author} {\bibfnamefont
  {A.}~\bibnamefont {Tsuda}}, \ and\ \bibinfo {author} {\bibfnamefont
  {A.}~\bibnamefont {Osuka}},\ }\bibfield  {title} {\enquote {\bibinfo {title}
  {Photophysical properties of porphyrin tapes},}\ }\href@noop {} {\bibfield
  {journal} {\bibinfo  {journal} {Journal of the American Chemical Society}\
  }\textbf {\bibinfo {volume} {124}},\ \bibinfo {pages} {14642--14654}
  (\bibinfo {year} {2002})}\BibitemShut {NoStop}%
\bibitem [{\citenamefont {Tittelbach-Helmrich}\ and\ \citenamefont
  {Steer}(1995)}]{tittelbach1995measurements}%
  \BibitemOpen
  \bibfield  {author} {\bibinfo {author} {\bibfnamefont {D.}~\bibnamefont
  {Tittelbach-Helmrich}}\ and\ \bibinfo {author} {\bibfnamefont {R.~P.}\
  \bibnamefont {Steer}},\ }\bibfield  {title} {\enquote {\bibinfo {title}
  {Measurements of the subpicosecond relaxation rates of the first excited
  singlet states of some pseudoazulenes in solution},}\ }\href@noop {}
  {\bibfield  {journal} {\bibinfo  {journal} {Chemical physics}\ }\textbf
  {\bibinfo {volume} {197}},\ \bibinfo {pages} {99--106} (\bibinfo {year}
  {1995})}\BibitemShut {NoStop}%
\bibitem [{Note1()}]{Note1}%
  \BibitemOpen
  \bibinfo {note} {Calculation of rates is based on the following
  approximations: The $S_0$ and $S_1$ potential energy surfaces are harmonic,
  share the same normal modes and frequencies but have different equilibrium
  geometries (shifted harmonic oscillators). Frequencies and normal modes are
  determined from a frequency calculation using $\omega $B97XD/dev2-SVG at the
  $S_0$ minimum. The Huang-Rhys factors are deduced from the gradient on $S_1$
  at the Franck-Condon point. Total rates are obtained by summing over all
  transitions that start in the initial vibrational ground state on $S_1$ and
  lead to a final vibrational state on $S_0$. Radiative transitions may lead to
  any vibrational state lower in energy, while in non-radiative transitions the
  final vibrational states on $S_0$ have to be approximately isoenergetic with
  the initial vibrational state on $S_1$. Following Ref. \cite
  {santoro2007effective} final states are grouped into classes C$_n$ depending
  on the number $n$ of simultaneously excited modes. The summation is limited
  to classes $C_0-C_8$, which captures most of the radiative rate and a large
  fraction of the non-radiative rate. Modes are sorted in decreasing order by
  Franck-Condon factors and the number of modes from which excitations are
  allowed, is reduced, until there are no more than $10^8$ elements per class
  left.}\BibitemShut {Stop}%
\bibitem [{\citenamefont {Budiardja}\ and\ \citenamefont
  {Cardall}(2011)}]{budiardja2011parallel}%
  \BibitemOpen
  \bibfield  {author} {\bibinfo {author} {\bibfnamefont {R.~D.}\ \bibnamefont
  {Budiardja}}\ and\ \bibinfo {author} {\bibfnamefont {C.~Y.}\ \bibnamefont
  {Cardall}},\ }\bibfield  {title} {\enquote {\bibinfo {title} {Parallel
  fft-based poisson solver for isolated three-dimensional systems},}\
  }\href@noop {} {\bibfield  {journal} {\bibinfo  {journal} {Computer Physics
  Communications}\ }\textbf {\bibinfo {volume} {182}},\ \bibinfo {pages}
  {2265--2275} (\bibinfo {year} {2011})}\BibitemShut {NoStop}%
\bibitem [{\citenamefont {Breneman}\ and\ \citenamefont
  {Wiberg}(1990)}]{breneman1990determining}%
  \BibitemOpen
  \bibfield  {author} {\bibinfo {author} {\bibfnamefont {C.~M.}\ \bibnamefont
  {Breneman}}\ and\ \bibinfo {author} {\bibfnamefont {K.~B.}\ \bibnamefont
  {Wiberg}},\ }\bibfield  {title} {\enquote {\bibinfo {title} {Determining
  atom-centered monopoles from molecular electrostatic potentials. the need for
  high sampling density in formamide conformational analysis},}\ }\href@noop {}
  {\bibfield  {journal} {\bibinfo  {journal} {Journal of Computational
  Chemistry}\ }\textbf {\bibinfo {volume} {11}},\ \bibinfo {pages} {361--373}
  (\bibinfo {year} {1990})}\BibitemShut {NoStop}%
\bibitem [{\citenamefont {Heringer}\ \emph {et~al.}(2007)\citenamefont
  {Heringer}, \citenamefont {Niehaus}, \citenamefont {Wanko},\ and\
  \citenamefont {Frauenheim}}]{heringer2007analytical}%
  \BibitemOpen
  \bibfield  {author} {\bibinfo {author} {\bibfnamefont {D.}~\bibnamefont
  {Heringer}}, \bibinfo {author} {\bibfnamefont {T.~A.}\ \bibnamefont
  {Niehaus}}, \bibinfo {author} {\bibfnamefont {M.}~\bibnamefont {Wanko}}, \
  and\ \bibinfo {author} {\bibfnamefont {T.}~\bibnamefont {Frauenheim}},\
  }\bibfield  {title} {\enquote {\bibinfo {title} {Analytical excited state
  forces for the time-dependent density-functional tight-binding method},}\
  }\href@noop {} {\bibfield  {journal} {\bibinfo  {journal} {Journal of
  computational chemistry}\ }\textbf {\bibinfo {volume} {28}},\ \bibinfo
  {pages} {2589--2601} (\bibinfo {year} {2007})}\BibitemShut {NoStop}%
\bibitem [{\citenamefont {Santoro}\ \emph {et~al.}(2007)\citenamefont
  {Santoro}, \citenamefont {Improta}, \citenamefont {Lami}, \citenamefont
  {Bloino},\ and\ \citenamefont {Barone}}]{santoro2007effective}%
  \BibitemOpen
  \bibfield  {author} {\bibinfo {author} {\bibfnamefont {F.}~\bibnamefont
  {Santoro}}, \bibinfo {author} {\bibfnamefont {R.}~\bibnamefont {Improta}},
  \bibinfo {author} {\bibfnamefont {A.}~\bibnamefont {Lami}}, \bibinfo {author}
  {\bibfnamefont {J.}~\bibnamefont {Bloino}}, \ and\ \bibinfo {author}
  {\bibfnamefont {V.}~\bibnamefont {Barone}},\ }\bibfield  {title} {\enquote
  {\bibinfo {title} {Effective method to compute franck-condon integrals for
  optical spectra of large molecules in solution},}\ }\href@noop {} {\bibfield
  {journal} {\bibinfo  {journal} {The Journal of chemical physics}\ }\textbf
  {\bibinfo {volume} {126}},\ \bibinfo {pages} {084509} (\bibinfo {year}
  {2007})}\BibitemShut {NoStop}%
\end{thebibliography}%

\end{document}